\documentclass{jfm}

\usepackage{amsmath}
\usepackage{xcolor}

\usepackage{natbib}
\usepackage{hyperref}
\usepackage{overpic}
\usepackage{color,soul}
\hypersetup{
    colorlinks = true,
    urlcolor   = blue,
    citecolor  = blue,
}

\newcommand{\RomanNumeralCaps}[1]
\linenumbers

\usepackage{url}
\usepackage{framed}
\colorlet{shadecolor}{yellow}

\newcommand{\lap}{\nabla^2}

\newcommand{\ez}{\boldsymbol{e}_{z}}

\newcommand{\rbase}{r}
\newcommand{\rvec}{\boldsymbol{\rbase}}
\newcommand{\velbase}{u}
\newcommand{\Velbase}{U}
\newcommand{\press}{p}
\renewcommand{\v}{\boldsymbol{\velbase}}
\newcommand{\vw}{\v_{\w}}
\newcommand{\dvw}[2]{\mathrm{d}\vw(#1,#2)}
\newcommand{\Vinf}{\boldsymbol{\Velbase}_{\infty}}
\newcommand{\vortbase}{\omega}
\newcommand{\w}{\boldsymbol{\vortbase}}
\newcommand{\spot}{\varphi}
\newcommand{\green}{G}

\newcommand{\vorkernbase}{P}
\newcommand{\vorkern}[2]{\vorkernbase(#1-#2)}
\newcommand{\regvorkern}[3]{\vorkernbase_{#3}(#1-#2)}

\newcommand{\vintkernbase}{\Pi}
\newcommand{\vintkern}[3]{\vintkernbase(#1-#2,#1-#3)}
\newcommand{\regvintkern}[4]{\vintkernbase_{#4}(#1-#2,#1-#3)}

\newcommand{\matkernbase}{\mathsf{P}}
\newcommand{\matkern}{\mathsfbi{P}}

\newcommand{\Gamvec}{\boldsymbol{\Gamma}}

\newcommand{\nv}{N}
\newcommand{\vjdex}{\mathsf{J}}
\newcommand{\vkdex}{\mathsf{K}}
\newcommand{\videx}{\mathsf{I}}

\newcommand{\blobrad}{\epsilon}

\newcommand{\axisjk}{\boldsymbol{\tau}_{\vjdex\vkdex}}

\newcommand{\djk}{d_{\vjdex\vkdex}}
\newcommand{\rvecscale}{\tilde{\rvec}}
\newcommand{\rvecjk}{\overline{\rvec}_{\vjdex\vkdex}}

\newcommand{\dvol}{\mathrm{d}V}

\newcommand{\state}{\mathsf{x}}
\newcommand{\joint}{\mathsf{z}}
\newcommand{\State}{\mathsf{X}}
\newcommand{\covar}{\Sigma}
\newcommand{\precis}{\covar^{-1}}

\newcommand{\meanstate}{\overline{\state}}

\newcommand{\approxstatecovar}{\covar^{\ast}_{\State}}

\newcommand{\expect}{E}

\newcommand{\gmmwt}{\alpha}
\newcommand{\gmmmeanbase}{\overline{\state}}
\newcommand{\gmmcovarbase}{\covar}
\newcommand{\gmmmean}[1]{\gmmmeanbase^{(#1)}}
\newcommand{\gmmcovar}[1]{\gmmcovarbase^{(#1)}}

\newcommand{\meanG}{\overline{\Gamma}}

\newcommand{\Reals}{\mathbb{R}}
\newcommand{\statedim}{n}
\newcommand{\measdim}{d}
\newcommand{\uniformdist}[1]{\mathcal{U}_{#1}}
\newcommand{\normaldist}{\mathcal{N}}
\newcommand{\probdist}{\pi}
\newcommand{\approxprob}{\hat{\probdist}}
\newcommand{\like}{L}
\newcommand{\logprob}{l}
\newcommand{\bounds}{B}
\newcommand{\constraint}{c_{\bounds}}

\newcommand{\sensbase}{s}

\newcommand{\rsensa}[1]{\boldsymbol{\sensbase}_{#1}}

\newcommand{\sensdex}{\alpha}

\newcommand{\observe}{h}
\newcommand{\observemat}{H}
\newcommand{\meas}{\mathsf{y}}
\newcommand{\Meas}{\mathsf{Y}}

\newcommand{\truemeas}{\meas^{\ast}}
\newcommand{\truestate}{\state^{\ast}}
\newcommand{\noise}{\varepsilon}
\newcommand{\Noise}{\mathcal{E}}
\newcommand{\noisecovar}{\covar_{\Noise}}

\newcommand{\noisestddev}{\sigma_{\Noise}}
\newcommand{\noisevar}{\noisestddev^{2}}
\newcommand{\ident}{I}
\newcommand{\svbase}{s}
\newcommand{\sv}[1]{\svbase_{#1}}
\newcommand{\svmat}{S}
\newcommand{\svsquaremat}{D}
\newcommand{\svsquare}[1]{\svbase^{2}_{#1}}

\newcommand{\eigenSx}{\lambda}
\newcommand{\Hrank}{r}

\newcommand{\rightvec}{v}

\newcommand{\change}[1]{\textcolor{black}{#1}}

\newcommand{\sect}{Section}

\title{Bayesian inference of vorticity in unbounded \\ flow from limited pressure measurements}

\author{Jeff D. Eldredge \corresp{\email{jdeldre@ucla.edu}} 
\and Mathieu Le Provost \corresp{Current address: Post-doctoral researcher, Department of Aeronautics \& Astronautics, Massachusetts Institute of Technology}}

\affiliation{Mechanical and Aerospace Engineering, University of California, Los Angeles\\ Los Angeles, CA 90095-1597 USA}
\begin{document}
\maketitle

\begin{abstract}
We study the instantaneous inference of an unbounded planar flow from sparse noisy pressure measurements. The true flow field comprises one or more regularized point vortices of various strength and size. We interpret the true flow's measurements with a vortex estimator, also consisting of regularized vortices, and attempt to infer the positions and strengths of this estimator assuming little prior knowledge. The problem often has several possible solutions, many due to a variety of symmetries. To deal with this ill-posedness and to quantify the uncertainty, we develop the vortex estimator in a Bayesian setting. We use Markov-chain Monte Carlo and a Gaussian mixture model to sample and categorize the probable vortex states in the posterior distribution, tailoring the prior to avoid spurious solutions. Through experiments with one or more true vortices, we reveal many aspects of the vortex inference problem. With fewer sensors than states, the estimator infers a manifold of equally-possible states. Using one more sensor than states ensures that no cases of rank deficiency arise. Uncertainty grows rapidly with distance when a vortex lies outside of the vicinity of the sensors. Vortex size cannot be reliably inferred, but the position and strength of a larger vortex can be estimated with a much smaller one. In estimates of multiple vortices their individual signs are discernible because of the non-linear coupling in the pressure. When the true vortex state is inferred from an estimator of fewer vortices, the estimate approximately aggregates the true vortices where possible.
\end{abstract}

\section{Introduction}
\label{sec:intro}

In a wide variety of practical situations, we wish to infer the state of a fluid flow from a limited number of flow sensors \change{with generally noisy output signals}. In particular, such knowledge of the flow state may assist within the larger scope of a flow control task, either in the training or application of a control strategy. For example, in reinforcement learning (RL) applications for guiding a vehicle to some target through a highly-disturbed fluid environment \citep{Verma2018}, the system is partially observable if the RL framework only has knowledge of the state of the vehicle itself. It is unable to distinguish between quiescent and disturbed areas of the environment and take actions that are distinctly advantageous in either, thus limiting the effectiveness of the control strategy. Augmenting this state with some knowledge of the flow may be helpful to improve this effectiveness.

The problem of flow estimation is very broad and can be pursued with different types of sensors in the presence of various flow physics. We focus in this paper on the inference of incompressible flows of moderate and large Reynolds numbers from pressure sensors, a problem that has been of \change{interest in the fluid dynamics community for many years \citep{Naguib2001,Murray2003,Gomez2019,Sashittal2021,Iacobello2022,Zhong2023}}. Flow estimation from other types of noisy measurements has also been pursued in closely-related contexts in recent years, with tools very similar to those used in the present work \citep{Juniper2022,Kontogiannis2022}. Estimation generally seeks to infer a finite-dimensional state vector $\state$ from a finite-dimensional measurement vector $\meas$. Since the state of a fluid flow is inherently infinite-dimensional, a central task of flow estimation is to approximately represent the flow so that it can be parameterized by a finite-dimensional state vector. For example, this could be done by a linear decomposition into data-driven modes \change{(e.g., with Proper Orthogonal Decomposition (POD) or Dynamic Mode Decomposition (DMD))}---in which the flow state comprises the coefficients of these modes---or by a generalized (non-linear) form of this decomposition via a \change{neural network \citep{Morimoto2022,fukai2023nature}}. Though these are very effective in representing flows close to their training set, they are generally less effective in representing newly-encountered flows. For example, a vehicle's interaction with a gust (an incident disturbance) may take very different forms, but the basis modes or neural network can only be trained on a subset of these interactions. Furthermore, even when these approaches are effective, it is difficult to probe them for intuition about some basic questions that underlie the estimation task. How does the effectiveness of the estimation depend on the physical distance between the sensors and vortical flow structures, or on the size of the structures?

We take a different approach to flow representation in this paper, writing the vorticity field as a sum of $\nv$ (nearly-)singular vortex elements. (The adjective ``nearly'' conveys that we will regularize each element with a smoothing kernel of small radius.) A distinct feature of a flow singularity (in contrast to, say, a POD mode) is that both its strength and its position are degrees of freedom in the state vector, so that it can efficiently and adaptively approximate a compact evolving vortex structure even with a small number of vortex elements. The compromise for this additional flexibility is that it introduces a non-linear relationship between the velocity field and the state vector. However, since pressure is already inherently quadratically dependent on the velocity, one must contend with non-linearity in the inference problem regardless of the choice of flow representation. Another advantage of using singularities as a representation of the flow is that their velocity and pressure fields are known exactly, providing useful insight into the estimation problem.

To restrict the dimensionality of the problem and make the estimation tractable, we truncate the set of vortex elements to a small number $\nv$. In this manner, the point vortices can be thought of as an adaptive low-rank representation of the flow, each capturing a vortex structure but omitting most details of the structure's internal dynamics. To keep the scope of this paper somewhat simpler, we will narrow our focus to unbounded two-dimensional vortical flows, so that the estimated vorticity field is given by
\begin{equation}
\omega(\rvec) = \sum_{\vjdex=1}^{\nv} \Gamma_{\vjdex} \tilde{\delta}_{\blobrad}(\rvec - \rvec_{\vjdex}),
\label{eq:sing-vort}
\end{equation}
where $\tilde{\delta}_{\blobrad}$ is a regularized form of the two-dimensional Dirac delta function with small radius $\blobrad$, and any vortex $\vjdex$ has strength $\Gamma_{\vjdex}$ and position described by two Cartesian coordinates $\rvec_{\vjdex} = (x_{\vjdex},y_{\vjdex})$. As we show in Appendix~\ref{sec:unbound}, the pressure field due to a set of $\nv$ vortex elements is given by
\begin{equation}
\press(\rvec) - \press_{\infty} = -\frac{1}{2}\rho \sum_{\vjdex=1}^{\nv} \Gamma^{2}_{\vjdex} \regvorkern{\rvec}{\rvec_{\vjdex}}{\blobrad} - \rho \sum_{\vjdex=1}^{\nv} \sum_{\vkdex = 1}^{\vjdex-1} \Gamma_{\vjdex}\Gamma_{\vkdex} \regvintkern{\rvec}{\rvec_{\vjdex}}{\rvec_{\vkdex}}{\blobrad},
\label{eq:pvpfull0}
\end{equation}
where $\vorkernbase_{\blobrad}$ is a regularized direct vortex kernel, representing the individual effect of a vortex on the pressure field, and $\vintkernbase_{\blobrad}$ is a regularized vortex interaction kernel, encapsulating the coupled effect of a pair of vortices on pressure; their details are provided in Appendix~\ref{sec:unbound}. This focus on unbounded two-dimensional flows preserves the essential purpose of the study, to reveal the important aspects of vortex estimation from pressure, and postpones to a future paper the effect of a body's presence or other flow contributors. Thus, the state dimensionality of the problems in this paper will be $\statedim = 3\nv$, composed of the positions and strengths of the $\nv$ vortex elements.

In this limited context, we address the following question: Given a set of noisy pressure measurements at $\measdim$ observation points (sensors) in or adjacent to an incompressible two-dimensional flow, to what extent can we infer a distribution of vortices? It is important to make a few points before we embark on our answer. First, because of the noise in measurements, we will address the inference problem in a probabilistic (i.e., Bayesian) manner: find the distribution of probable states based on the likelihood of the true observations. As we have noted, we have no hope of approximating a smoothly-distributed vorticity field with a small number of singular vortex elements. However, as a result of our probabilistic approach, the {\em expectation} of the vorticity field over the whole distribution of estimated states will be smooth, even though the vorticity of each realization of the estimated flow is singular in space. This fact is shown in Appendix~\ref{sec:expect-vort}.

Second, vortex flows are generally unsteady, so we ultimately wish to address this inference question over some time interval. Indeed, that has been the subject of several previous works, e.g., \cite{darakjdeprf18,provost2021ensemble,leprovost2022lrenkf}, in which pressure measurements were assimilated into the estimate of the state via an ensemble Kalman filter (EnKF) \citep{Evensen1994}. Each step of such sequential data assimilation consists of the same Bayesian inference (or {\em analysis}) procedure: we start with an initial guess for the probability distribution (the prior) and seek an improved guess (the posterior). At steps beyond the first one, we generate this prior by simply advancing an ensemble of vortex element systems forward by one time step (the {\em forecast}), from states drawn from the posterior at the end of the previous step. The quality of the estimation generally improves over time as the sequential estimator proceeds. However, when we start the problem we have no such posterior from earlier steps. \change{Furthermore, even at later times, as new vortices are created or enter the region of interest, the prior will lack a description of these new features.}

\change{This challenge forms a central task} of the present paper: We seek to infer the flow state at some instant from a prior that expresses little to no knowledge of the flow. Aside from some loose bounds, we do not have any guess for where the vortex elements lie, how strong they are, or even how many there should be. All we know of the true system's behavior comes from the sensor measurements, and we therefore estimate the vortex state by maximizing the likelihood that these sensor measurements will arise. It should also be noted that viscosity has no effect on the instantaneous relationships between vorticity, velocity, and pressure in unbounded flow, so it is irrelevant if the true system is viscous or not.  To assess the success of our inference approach, we will compute the expectation of the vorticity field under our estimated probability distribution and compare it with the true field, as we will presume to know this latter field for testing purposes.  



The last point to make is that the inference of a flow from pressure is often an ill-posed problem with multiple possible solutions, a common issue with inverse problems of partial differential equations. For example, we will find that there may be many candidate vortex systems that reasonably match the pressure sensor data, forming ridges or local maxima of likelihood, even if they are not the global maximum solution. As we will show, these situations arise most frequently when the number of sensors is less than or equal to the number of states, i.e., when the inverse problem is underdetermined to some degree. In these cases, we will find that adding even one additional sensor can address the underlying ill-posedness. \change{There will also be various symmetries that arise due to the vortex-sensor arrangement.} In this paper, we will use techniques to mitigate the effect of these symmetries on the estimation task. However, multiple solutions may still arise even with such techniques, and we seek to explore this multiplicity thoroughly. Therefore, we adopt a solution strategy that explores all features of the likelihood function, including multiple maxima. We describe the probability-based formulation of the vortex estimation problem and our methodologies for revealing it in Section~\ref{sec:problem}. Then, we present results of various estimation exercises with this methodology in Section~\ref{sec:results}\change{, and discuss the results and their generality in Section~\ref{sec:conclude}.}


\section{Problem statement and methodology}
\label{sec:problem}

\subsection{The inference problem}
\label{sec:inferenceproblem}

Our goal is to estimate the state of an unbounded two-dimensional vortical flow with vortex system (\ref{eq:sing-vort}), which we will call a {\em vortex estimator}, specified completely by the $\statedim$-dimensional vector ($\statedim=3\nv$)
\begin{equation}
\state = (\state_{1},\ldots,\state_{\nv})^{T} \in \Reals^{3\nv},
\label{eq:stateblock}
\end{equation}
where
\begin{equation}
\state_{\vjdex} \equiv (\rvec_{\vjdex},\Gamma_{\vjdex}) = (x_{\vjdex},y_{\vjdex},\Gamma_{\vjdex})
\label{eq:stateblockj}
\end{equation}
is the 3-component state of vortex $\vjdex$. The associated state covariance matrix is written as
\begin{equation}
\covar_{\state} = \begin{pmatrix} \covar_{11} &\covar_{12} &\cdots &\covar_{1\nv} \\ \covar_{21} & \covar_{22} & \cdots & \covar_{2\nv} \\ \vdots & \vdots & \ddots & \vdots \\ \covar_{\nv 1} & \covar_{\nv 2} &\cdots & \covar_{\nv\nv}\end{pmatrix} \in \Reals^{3\nv\times3\nv},
\label{eq:covarblock}
\end{equation}
where each $3\times 3$ block represents the covariance between vortex elements $\vjdex$ and $\vkdex$:
\begin{equation}
\covar_{\vjdex\vkdex} \equiv  \begin{pmatrix} \covar_{\rvec_{\vjdex} \rvec_{\vkdex}} & \covar_{\rvec_{\vjdex} \Gamma_{\vkdex}} \\
\covar_{\Gamma_{\vjdex} \rvec_{\vkdex}} & \covar_{\Gamma_{\vjdex} \Gamma_{\vkdex}} \end{pmatrix}.
\label{eq:covarblockjk}
\end{equation}
Note that $\covar_{\vkdex\vjdex} = \covar^{T}_{\vjdex\vkdex}$.

Equation \eqref{eq:pvpfull0} expresses the pressure (relative to ambient), $\Delta\press(\rvec) \equiv \press(\rvec) - \press_{\infty}$, as a continuous function of space, $\rvec$. Here, we also explicitly acknowledge its dependence on the state, $\Delta\press(\rvec,\state)$. Furthermore, we will limit our observations to a finite number of sensor locations, $\rvec = \rsensa{\sensdex}$, for $\sensdex = 1,\ldots,\measdim$, and define from this an {\em observation operator}, $\observe: \Reals^{\statedim} \rightarrow \Reals^{\measdim}$, mapping a given state vector $\state$ to the pressure at these sensor locations:
\begin{equation}
\observe(\state) \equiv\left( \Delta\press(\rsensa{1},\state), \Delta\press(\rsensa{2},\state),\ldots,\Delta\press(\rsensa{\measdim},\state)\right)^{T} \in \Reals^{d}.
\label{eq:pobserve}
\end{equation}
The objective of this paper is essentially to explore the extent to which we can invert function (\ref{eq:pobserve}): from a given set of pressure observations $\truemeas \in \Reals^{\measdim}$ of the true system at $\rsensa{\sensdex}$, $\sensdex = 1,\ldots,\measdim$, determine the state $\state$. \change{In this work, the true sensor measurements $\truemeas$ (the truth data) will be synthetically generated from the pressure field of a set of vortex elements in unbounded quiescent fluid, obtained from the same expression for pressure \eqref{eq:pvpfull0} that we use for $\observe(\state)$ in the estimator. Throughout, we will refer to the set of vortices that generated the measurements as the true vortices.} However, there is inherent uncertainty in $\truemeas$ due to random measurement noise $\noise \in \Reals^{\measdim}$, so we model the predicted observations $\meas$ as 
\begin{equation}
 \meas = \observe(\state) + \noise,
 \label{eq:noisyobs}
 \end{equation}
\change{where $\noise$ is normally distributed about zero mean, $\normaldist(\noise| 0,\noisecovar)$, and the sensor noise is assumed independent and identically distributed, so that its covariance is $\noisecovar = \noisevar\ident$ with $\ident \in \Reals^{\measdim\times\measdim}$ the identity. We seek a probabilistic form of the inversion of (\ref{eq:noisyobs}) when set equal to $\truemeas$.} That is, we seek the conditional probability distribution of states based on the observed data, $\probdist(\state|\truemeas)$: the peaks of this distribution would represent the most probable state(s) based on the measurements, and the breadth of the peaks would represent our uncertainty about the answer.

From Bayes' theorem, the conditional probability of the state given an observation, $\probdist(\state|\meas)$, can be regarded as a posterior distribution \change{over $\state$},
\begin{equation}
\probdist(\state|\meas) = \frac{\like(\meas|\state) \probdist_{0}(\state)}{\probdist(\meas)},
\label{eq:bayes}
\end{equation} 
where $\probdist_{0}(\state)$ is the prior distribution, describing our original beliefs about the state $\state$, and $\like(\meas|\state)$ is called the likelihood function, representing the probability of observing certain data $\meas$ at a given state, $\state$. The likelihood function encapsulates our physics-based prediction of the sensor measurements, based on the observation operator $\observe(\state)$. Collectively, $\like(\meas|\state) \probdist_{0}(\state)$ represents the joint distribution of states and their associated observations. The distribution of observations, $\probdist(\meas)$, \change{is a uniform normalizing factor. Its value is unnecessary for characterizing the posterior distribution over $\state$, since only comparisons (ratios) of the posterior are needed during sampling, as we discuss below in Section~\ref{sec:sample}. Thus, we can omit the denominator in (\ref{eq:bayes}).} We evaluate this unnormalized posterior at the true observations, $\truemeas$, and denote it by $\tilde{\probdist}(\state|\truemeas) = \like(\truemeas|\state) \probdist_{0}(\state)$.

Our goal is to explore and characterize this unnormalized posterior for the vortex system. \change{Expressing our lack of prior knowledge,} we write $\probdist_{0}(\state)$ as a uniform distribution within a certain acceptable bounding region $\bounds$ on the state components \change{(discussed in more specific detail below)},
\begin{equation}
\probdist_{0}(\state) = \uniformdist{\statedim}(\state|\bounds).
\end{equation}
Following from our observation model \eqref{eq:noisyobs} \change{with Gaussian noise}, the likelihood is a Gaussian distribution about the predicted observations, $\observe(\state)$:
\begin{equation}
\like(\meas|\state) = \normaldist(\meas|\observe(\state),\noisecovar) = \frac{1}{\sqrt{(2\pi)^{\measdim}\det \noisecovar}} \exp\left(-\frac{1}{2} ||\meas - \observe(\state)||^{2}_{\noisecovar}\right),
\label{eq:likenormal}
\end{equation}
where we have defined the covariance-weighted norm
\begin{equation}
||\meas||^{2}_{\covar} \equiv \meas^{T}\precis\meas.
\end{equation}
Thus, our unnormalized posterior for the vortex estimator is given by
\begin{equation}
\tilde{\probdist}(\state|\truemeas) = \frac{\uniformdist{\statedim}(\state|\bounds)}{\sqrt{(2\pi)^{\measdim}\det \noisecovar}} \exp\left(-\frac{1}{2} ||\truemeas - \observe(\state)||^{2}_{\noisecovar}\right).
\label{eq:unscalepost}
\end{equation}
For practical purposes it is helpful to take the log of this probability, so that ratios of probabilities---some of them near machine zero---are assessed instead via differences in their logs. \change{Because only differences are relevant, we can dispense with constants that arise from taking the log, such as the inverse square root factor. We note that the uniform probability distribution $\uniformdist{\statedim}(\state|\bounds)$ is uniform and positive inside $\bounds$ and zero outside.} Thus, to within an additive constant, this log-posterior is
\begin{equation}
\logprob(\state|\truemeas) = - \change{\frac{1}{2}} ||\truemeas - \observe(\state)||^{2}_{\noisecovar} + \constraint(\state),
\label{eq:logprob}
\end{equation}
where $\constraint(\state)$ is a barrier function \change{arising from the log of the uniform distribution}, equal to zero for any $\state$ inside of the restricted region $\bounds$ of our uniform distribution, and $\constraint(\state)= -\infty$ outside of $\bounds$.

\change{In the examples that we present in this paper, the pressure sensors will be uniformly distributed along a straight line on the $x$ axis, unless otherwise specified. and the set of vortices used for estimation purposes as the vortex estimator. To ensure finite pressures throughout the domain, both the true vortices and the vortex estimator are regularized as discussed in Appendix~\ref{sec:regularized}, with a small blob radius $\blobrad = 0.01$ unless otherwise stated. To improve the scaling and conditioning of the problem, the pressure (relative to ambient) is implicitly normalized by $\rho \Gamma_{0}^{2}/L^{2}$, where $\Gamma_{0}$ is the strength of the largest-magnitude vortex in the true set and $L$ represents a characteristic distance of the vortex set from the sensors; all positions are implicitly normalized by $L$ and vortex strengths by $\Gamma_{0}$. Unless otherwise specified, the measurement noise is $\noisestddev = 5\times 10^{-4}$.}

\change{\subsection{Symmetries and non-linearity in the vortex-pressure system}}

\begin{figure}
\begin{center}
\includegraphics{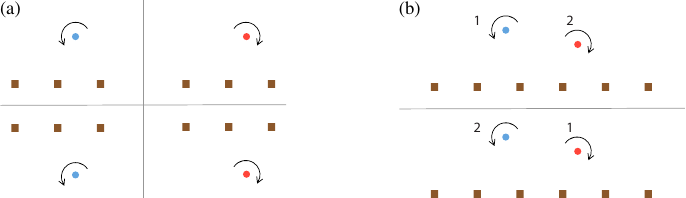}
\end{center}
\caption{(a) Four configurations of vortices that would generate identical measurements for the set of pressure sensors (brown squares). (b) Two distinct vortex states that differ only in the vortex labeling but generate identical flow fields.}\label{fig:vortex-schematic}
\end{figure}

\change{As mentioned in the introduction, there are many situations in which multiple solutions arise due to symmetries. This is easy to see from a simple thought experiment, depicted in Figure~\ref{fig:vortex-schematic}(a). Suppose that we wish to estimate a single vortex from pressure sensors arranged in a straight line. A vortex on either side of this line of sensors will induce the same pressure on the sensors, and a vortex of either sign of strength will, as well. Thus, in this simple problem, there are four possible states that are indistinguishable from each other, and we would need more information about the circumstances of the problem to rule out three of them. Such symmetries arise commonly in the problems we will study in this paper.}

\begin{figure}
\begin{center}
\includegraphics[width=0.78\textwidth]{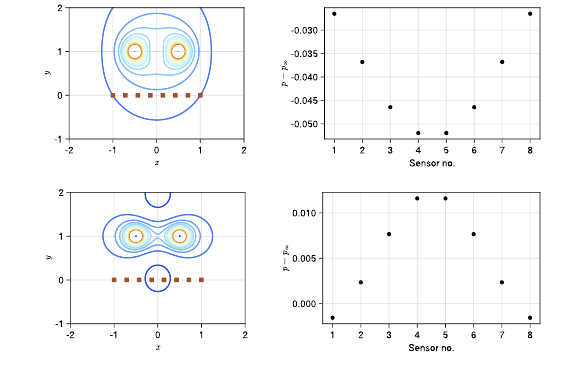}
\end{center}
\caption{Pressure fields and associated sensor measurements for two examples of a pair of vortex elements at $\rvec_{1} = (-1/2,1)$ and $\rvec_{2} = (1/2,1)$, respectively. Sensors are depicted in the pressure field plots as brown squares. Top: $\Gamma_{1} = 1$ and $\Gamma_{2} = 1$. Bottom: $\Gamma_{1} = 1$ and $\Gamma_{2} = -1$. In both cases, pressure contours are between $-0.5$ (red) and $0.01$ (blue).}\label{fig:2vortex-pressure}
\end{figure}

\change{The symmetry with respect to the sign of vortex strength is due to the non-linear relationship between pressure and vorticity. However, it is important to note that this symmetry issue is partly alleviated by the non-linear relationship, as well, because of the coupling that it introduces between vortices. Figure~\ref{fig:2vortex-pressure} depicts the pressure fields for two examples of a pair of vortices: one in which the vortices in the pair have equal strengths and another in which the vortices have equal but opposite strengths. Though the pressure in the vortex cores of both pairs is similar and sharply negative, the pressures outside the cores are distinctly different because the interaction kernel enters the sum with different sign. At the positions of the sensors, the pressure has a small region of positive pressure in the case of vortices of opposite sign. These differences are essential for inferring the relative signs of vortex strengths in sets of vortices. However, it is important to stress that the pressure is invariant to a change of sign of all vortices in the set, so we would still need more prior information to discriminate one overall sign from the other.}

\change{Another symmetry arises when there is more than one vortex to estimate, as in Figure~\ref{fig:vortex-schematic}(b), because in such a case, there is no unique ordering of the vortices in the state vector. With each of the vortices assigned a fixed set of parameters, any of the $\nv!$ permutations of the ordering leads to the same pressure measurements. This vortex relabeling symmetry is a discrete analog of the particle relabeling symmetry in continuum mechanics \citep{marsden2013}; it is also closely analogous to the {\em non-identifiability} issue of the mixture models that will be used for the probabilistic modeling in this paper. All of the $\nv!$ solutions are obviously equivalent from a flow field perspective, so this symmetry is not a problem if we assess estimator performance based on flow field metrics. However, the $\nv!$ solutions form distinct points in the state space and we must anticipate this multiplicity when we search for high-probability regions.}

\change{The barrier function $\constraint(\state)$ in \eqref{eq:logprob} allows us to anticipate and eliminate some modes that arise from problem symmetries, because we can use the bounding region $B$ to reject samples that fail to meet certain criteria. To eliminate some of the aforementioned symmetries, we will, without loss of generality, restrict our vortex estimator to search for vortices that lie above the line of sensors on the $x$ axis. For cases of multiple vortices in the estimator, we re-order the vortex entries in the state vector by their $x$ position at each MCMC step to eliminate the relabeling symmetry. We also assume that the leftmost vortex has positive strength, which reduces the number of probable states by half; the signs of all estimated vortices can easily be switched a posteriori if new knowledge shows that this assumption is wrong. }

\subsection{The true covariance and rank deficiency}
\label{sec:gauss1}

\change{The main challenge of the vortex inference problem is that the observation operator is non-linear, so the posterior (\ref{eq:unscalepost}) is not Gaussian. Thus, we will instead sample this posterior and develop an approximate model for the samples, composed of a mixture of Gaussians. However, we can obtain some important insight by supposing that we already know the true state and then linearizing the observation operator about it,}
\begin{equation}
\observe(\state) \approx \observe(\truestate) + \observemat \cdot (\state-\truestate),
\label{eq:taylor}
\end{equation}
where $\observemat \equiv \nabla\observe(\truestate) \in \Reals^{\measdim\times\statedim}$, the Jacobian of the observation operator at the true state. Then we can derive an approximating $\statedim$-dimensional Gaussian model about mean $\truestate$ (plus a bias due to noise in the realization of the true measurements) with covariance
\begin{equation}
\approxstatecovar = (\observemat^{T}\noisecovar^{-1}\observemat)^{-1} \in \Reals^{\statedim\times\statedim}.
\label{eq:approxH1}
\end{equation}
A brief derivation of this result is included in Appendix~\ref{sec:bayes-gauss}. 

We will refer to $\approxstatecovar$ as the ``true'' state covariance. It is useful to note that the matrix $\observemat^{T}\noisecovar^{-1}\observemat$ is the so-called Fisher information matrix for our Gaussian likelihood \citep{Cui2021}, evaluated at the true state. In other words, it quantifies the information about the state of the system that is available in the measurements. Because all sensors have the same noise variance, $\noisevar$, then $\approxstatecovar = \noisevar(\observemat^{T}\observemat)^{-1}$. We can then use the singular value decomposition of the Jacobian, $\observemat = U \svmat V^{T}$, to write a diagonalized form of the covariance,
\begin{equation}
\approxstatecovar = V \Lambda V^{T}.
\label{eq:approxH2}
\end{equation}
Here, the eigenvalue matrix is $\Lambda = \noisevar \svsquaremat^{-1}$, where $\svsquaremat = \svmat^{T}\svmat \in \Reals^{\statedim\times\statedim}$ is a diagonal matrix containing squares $\svsquare{j}$ of the singular values $\svmat$ of $\observemat$ in decreasing magnitude up to the rank $\Hrank \leq \min(\measdim,\statedim)$ of $\observemat$ and padded with $\statedim - \Hrank$ zeros. The uncertainty ellipsoid thus has semi-axis lengths
\begin{equation}
\eigenSx_{j}^{1/2} = \noisestddev/\sv{j}
\label{eq:eigenSx}
\end{equation}
along the directions $\rightvec_{j}$ given by the corresponding columns of $V$. Thus, the greatest uncertainty $\eigenSx^{1/2}_{\statedim}$ is associated with the smallest singular value $\sv{\statedim}$ of $\observemat$. The corresponding eigenvector, $\rightvec_{\statedim}$, indicates the mixture of states for which we have the most confusion.

In fact, the smallest singular values of $\observemat$ are necessarily zero if $\statedim > \measdim$, i.e., when there are fewer sensors than states and the problem is therefore underdetermined. In such a case, $\observemat$ has a null space spanned by the last $\statedim-\Hrank$ columns in $V$. However, $\truestate$ is not a unique solution in these problems; rather, it is simply one element of a manifold of vortex states that produce equivalent sensor readings (to within the noise). The covariance $\approxstatecovar$ evaluated at any $\truestate$ on the manifold reveals the local tangent to this manifold---directions near $\truestate$ along which we get identical sensor values. The true covariance $\approxstatecovar$ will also be very useful for illuminating cases in which the problem is ostensibly fully determined ($\statedim \leq \measdim$), but for which the arrangement of sensors and true vortices nonetheless creates significant uncertainty in the estimate. In some of these cases, the smallest singular values may still be zero or quite small, indicating that the effective rank of $\observemat$ is smaller than $\statedim$.


\subsection{Sampling and modeling of the posterior}
\label{sec:sample}

\begin{figure}
\begin{center}
\includegraphics[width=\textwidth]{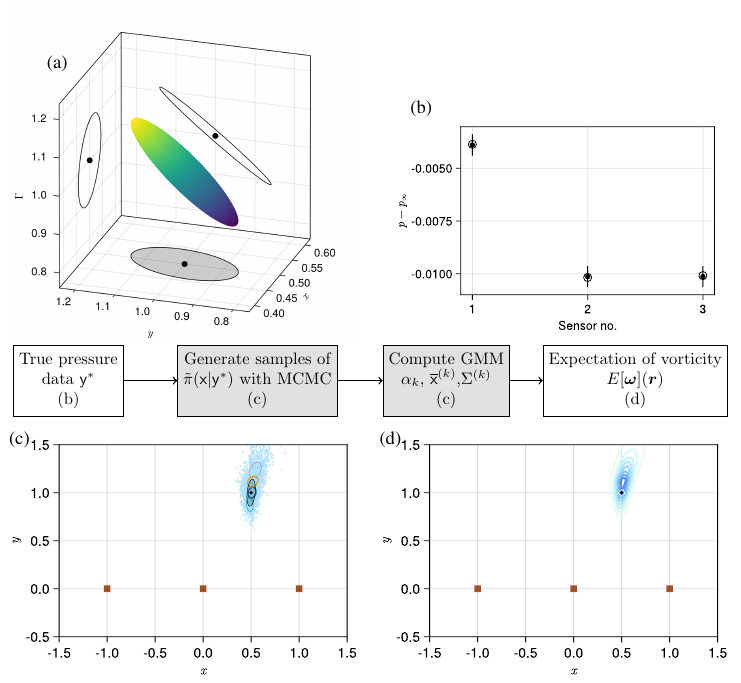}
\end{center}
\caption{One true vortex and one-vortex estimator using three sensors\change{, with center panel showing flowchart of overall algorithm and references to figure panels.} (a) True covariance ellipsoid. The ellipse on each coordinate plane represents the marginal covariance between the state components in that plane. The true vortex state is shown as a black dot. (b) True sensor data (filled circles) with noise levels (vertical lines), compared with sensor values from estimate (open circles), obtained from expected state of the sample set. (c) MCMC samples (blue) and the resulting vortex position covariance ellipses (colored unfilled ellipses) from the Gaussian mixture model. Thicker lines of the ellipses indicate higher weights in the mixture. The true vortex position is shown as a filled black circle, and the true covariance ellipse for vortex position (corresponding to the gray ellipse in (a)) is shown filled in gray. Sensor positions are shown as brown squares. (d) Contours of the expected vorticity field, based on the mixture model.}
\label{fig:onevortex-3sensor}
\end{figure}

\change{The true covariance matrix (\ref{eq:approxH1}) and its eigendecomposition will be an important tool in the study that follows, but we generally will only use it when we presume to know the true state $\truestate$ and seek illumination on the estimation \change{in the vicinity of the solution}. To characterize the problem more fully and explore the potential multiplicity of solutions, we will generate samples of the posterior and then fit the samples with an approximate distribution $\approxprob_{\truemeas}(\state) \approx \probdist(\state|\truemeas)$ over $\state$. The overall algorithm is shown in the center panel in Figure~\ref{fig:onevortex-3sensor} in the context of an example of estimating one vortex with three sensors. For the sampling task, we use the Metropolis--Hastings (MH) method \citep[see, e.g.,][]{Chib1995}, a simple but powerful form of Markov chain Monte Carlo (MCMC). This method relies only on differences of the log probabilities between a proposed chain entry and the current chain entry to determine whether the proposal is \change{is more probable than the previous entry and should be} added to the chain of samples. To ensure that the MCMC sampling does not get stuck in one of possibly multiple high-probability regions, we use the method of parallel tempering \citep{Sambridge2014}.}

\change{In practice, we generally have found good results in parallel tempering by using five parallel Markov chains exploring the target distribution raised to respective powers $3.5^{p}$, where $p$ takes integer values between $-4$ and $0$. We initially carry out $10^{4}$ steps of the algorithm with the MCMC proposal variance set to a diagonal matrix of $4\times 10^{-4}$ for every state component. Then, we perform $10^{6}$ steps with proposal variances set more tightly, uniformly equal to $2.5\times 10^{-5}$. The sample data set is then obtained from the $p = 0$ chain after discarding the first half of the chain (for burn-in) and retaining only every $100$th chain entry of the remainder to minimize autocorrelations in the samples. On a MacBook Pro with a M1 processor, the overall process takes around 20 seconds in the most challenging cases (i.e., the highest-dimensional state spaces). There are other methods that move more efficiently in the direction of local maxima (e.g., such as hybrid MCMC, which uses the Jacobian of the log-posterior to guide the ascent). However, the approach we have taken here is quite versatile for more general cases, particularly those in which the Jacobian is impractical to compute repetitively in a long sequential process.}

To approximate the posterior distribution over $\state$ from the samples, we employ a Gaussian mixture model (GMM), which assumes the form of a weighted sum of $K$ normal distributions (the mixture components),
\begin{equation}
\approxprob_{\truemeas}(\state) = \sum_{k=1}^{K}\gmmwt_{k} \normaldist(\state|\gmmmean{k},\gmmcovar{k}),
\label{eq:gmm}
\end{equation}
where $\gmmmean{k}$ and $\gmmcovar{k}$ are the mean and covariance of component $k$ of the mixture. Each weight $\gmmwt_{k}$ lies between 0 and 1 and represents the probability of a sample point ``belonging to'' component $k$, so it quantifies the importance of that component in the overall mixture. For a given set of samples and a pre-selected number of components $K$, the parameters of the GMM ($\gmmwt_{k}$, $\gmmmean{k}$, and $\gmmcovar{k}$) are found via the Expectation Maximization algorithm \citep{bishopbook}. It is generally advantageous to choose $K$ to be large ($\sim 5$--$9$) because extraneous components are assigned little weight, resulting in a smaller number of effective components. It should also be noted that, for the special case of a single mode of small variance, a GMM with $K=1$ and a sufficient number of samples approaches the Gaussian approximation about the true state $\truestate$ described in Section~\ref{sec:gauss1}, with covariance $\approxstatecovar$. 

As we show in Appendix~\ref{sec:expect-vort}, the GMM has a very attractive feature when used in the context of singular vortex elements, because we can exactly evaluate the expectation of the vorticity field under the GMM probability, 
\begin{equation}
\expect[\w](\rvec) = \sum_{k=1}^{K} \gmmwt_{k}\sum_{\vjdex=1}^{\nv} \left[ \meanG^{(k)}_{\vjdex} + \gmmcovar{k}_{\Gamma_{\vjdex} \rvec_{\vjdex}} {\gmmcovar{k}_{\rvec_{\vjdex} \rvec_{\vjdex}}}^{-1} (\rvec - \overline{\rvec}^{(k)}_{\vjdex} )\right] \normaldist(\rvec|\overline{\rvec}_{\vjdex}^{(k)},\covar^{(k)}_{\rvec_{\vjdex} \rvec_{\vjdex}}),
\label{eq:expected-vort}
\end{equation}
where $\overline{\rvec}^{(k)}_{\vjdex}$ and $\meanG^{(k)}_{\vjdex}$ comprise the position and strength of the $\vjdex$-vortex of mean state $\gmmmean{k}$ in equation~(\ref{eq:stateblock}), and $\gmmcovar{k}_{\rvec_{\vjdex}\rvec_{\vjdex}}$ and $\gmmcovar{k}_{\Gamma_{\vjdex} \rvec_{\vjdex}}$ are elements in the $\vjdex\vjdex$-block of covariance $\gmmcovar{k}$, defined in (\ref{eq:covarblockjk}). Thus, under a Gaussian mixture model of the state, the expected vorticity field is itself composed of a sum of Gaussian-distributed vortices in Euclidean space (due to the first term in the square brackets), plus a sum of Gaussian-regularized dipole fields arising from covariance between the strengths and positions of the inferred vortex elements (the second term in square brackets).

\section{Inference examples}
\label{sec:results}

\change{Though we have reduced the problem by non-dimensionalizing it and restricting the possible states, there remain several parameters to explore in the vortex estimation problem}: the number and relative configuration of the true vortices; the number of vortices used by the estimator; the number of sensors $\measdim$ and their configuration; and the measurement noise level $\noisestddev$.  We will also explore the inference of vortex radius $\blobrad$ when this parameter is included as part of the state. In the next section, we will explore the inference of a single vortex, using this example to investigate many of the parameters listed above. Then, in the following sections, we will examine the inference of multiple true vortices, to determine the unique aspects that emerge in this context.

\subsection{Inference of a single vortex}

In this section, we will explore cases in which both the true vortex system and our estimator consist of a single vortex. We will use this case to draw insight on many of the parameters of the inference problem. For most of the cases, a single line of sensors along the $x$ axis will be used. The true vortex will remain on the $y = 1$ line and have unit strength. Many of the examples that follow will focus on the true configuration $(x_{1},y_{1},\Gamma_{1}) = (0.5,1,1)$. (The subscript 1 is unnecessary with only one vortex, but allows us to align the notation with that of Section~\ref{sec:inferenceproblem}). Note, however, that we will not presume knowledge of any of these states: the bounding region of our prior will be $x \in (-2,2)$, $y \in (0.01,4)$, $\Gamma \in (0,2)$.

All of the basic tools used in the present investigation are depicted in Figure~\ref{fig:onevortex-3sensor}, in which the true configuration is estimated with three sensors arranged uniformly along the $x$ axis between $[-1,1]$ with $\noisestddev = 5\times 10^{-4}$. Figure~\ref{fig:onevortex-3sensor}(a) shows the ellipsoid for covariance $\approxstatecovar$, computed at the true vortex state. This figure particularly indicates that much of the uncertainty lies along a direction that mixes $y_{1}$ and $\Gamma_{1}$; indeed, the eigenvector corresponding to the direction of greatest uncertainty is $(0.08,0.79,0.61)$. This uncertainty is intuitive: as a vortex moves further away from the sensors, it generates very similar sensor measurements if its strength simultaneously increases in the proportion indicated by this direction. In Figure~\ref{fig:onevortex-3sensor}(b), the samples obtained from the MCMC method are shown. Here, and in later figures in the paper, we show only the vortex positions of these samples in Euclidean space and color their symbols to denote the sign of their strength (blue for positive, red for negative). The set of samples clearly encloses the true state, shown as a block dot; the expected value from the samples is $(0.51,1.07,1.05)$, which agrees well.

The samples also demonstrate the uncertainty of the estimated state. The filled ellipse in this figure corresponds to the exact covariance of Figure~\ref{fig:onevortex-3sensor}(a) and is shown for reference. As expected, the samples are spread predominantly along the direction of the maximum uncertainty. This figure also depicts an elliptical region for each Gaussian component of the mixture computed from the samples. These ellipses correspond only to the marginal covariances of the vortex positions and do not depict the uncertainty of the vortex strength. The weight of the component in the mixture is denoted by the thickness of the line. One can see from this plot that the GMM covers the samples with components, concentrating most of the weight near the center of the cluster with two dominant components. The composite of these components is best seen in Figure~\ref{fig:onevortex-3sensor}(d), in which the expected vorticity field is shown. In the remainder of this paper, this expected vorticity field will be used extensively to illuminate the uncertainty of the vortex estimation. Finally, Figure~\ref{fig:onevortex-3sensor}(c) compares the true sensor pressures with those corresponding to the expected state from the MCMC samples. These agree to within the measurement noise.

\begin{figure}
\begin{center}
\includegraphics[width=0.51\textwidth]{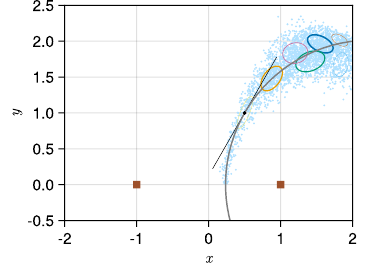}
\end{center}
\caption{One true vortex and one-vortex estimator, using two sensors, showing MCMC samples (blue dots) and the resulting vortex position ellipses from the Gaussian mixture model. The estimator truncates the samples outside the bounding region. The true vortex position is shown as filled black circle. The circular curve represents the manifold of possible states that produce the same sensor pressures; the line tangent to the circle is the direction of maximum uncertainty at the true state. }\label{fig:onevortex-samples-2sensor}
\end{figure}

\subsubsection{Effect of the number of sensors}

In the last section, we found that three sensors were sufficient to estimate a single vortex's position and strength. In this section we investigate how this estimate of a single vortex depends on the number of sensors. In most cases, these sensors will again lie uniformly along the $x$ axis in the range $[-1,1]$. Intuitively, we expect that if we have fewer sensors than there are states to estimate, we will have insufficient information to uniquely identify the vortex. Figure~\ref{fig:onevortex-samples-2sensor} shows that this is indeed the case. In this example, only two sensors are used to estimate the same vortex as in the previous example. The MCMC samples are distributed along a circular arc, but are truncated outside of the aforementioned bounding region. In fact, this arc is a projection of a helical curve of equally-probable states in the three-dimensional state space. The samples broaden from the arc the further they are from the sensors due to the increase in uncertainty with distance. The true covariance, $\approxstatecovar$, cannot reveal the full shape of this helical manifold, which is inherently dependent on the non-linear relationship between the sensors and the vortex. However, the rank of this covariance decreases to 2, so that the uncertainty along one of its principal axes must be infinite. This principal axis is tangent to the manifold of possible states, as shown by a line in the plot.

\begin{figure}
\begin{center}
\includegraphics[width=\textwidth]{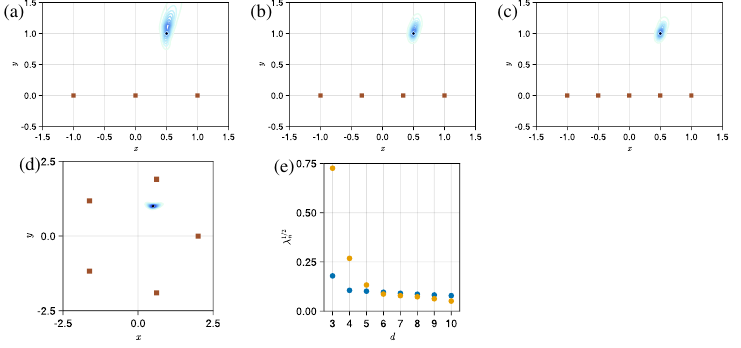}
\end{center}
\caption{One true vortex and one-vortex estimator, using various numbers and configurations of sensors (shown as brown squares) arranged in a line between -1 and 1 (a,b,c) or in a circle of radius 2.1 (d). Each panel depicts contours of expected vorticity field. True vortex position shown as filled black circle in each. (e) Maximum length of the covariance ellipsoid with increasing number of sensors on line segment $x = [-1,1]$ (blue) or circle of radius $2.1$ (gold).} \label{fig:1vortex-sensors}
\end{figure}

What if there are more sensors than states? Figure~\ref{fig:1vortex-sensors}(a,b,c) depict expected vorticity fields for several cases in which there are increasing numbers of sensors arranged along a line, and Figure~\ref{fig:1vortex-sensors}(d) shows the expected vorticity when 5 sensors are instead arranged in a circle of radius 2.1 about the true vortex. (The choice of radius is to ensure that the smallest distance between the true vortex and a sensor is approximately 1 in all cases.) It is apparent that the uncertainty shrinks when the number of sensors increases from 3 to 4, but does so less notably when the number increases from 4 to 5. In Figure~\ref{fig:1vortex-sensors}(e), the maximum uncertainty is seen to drop by nearly half when one sensor is added to the basic set of 3 along a line, but decreases much more gradually when more than 4 sensors are used. The drop in uncertainty is more dramatic between 3 and 5 sensors arranged in a circle, but becomes more gradual beyond 5 sensors.

%

\begin{figure}
\begin{center}
\includegraphics{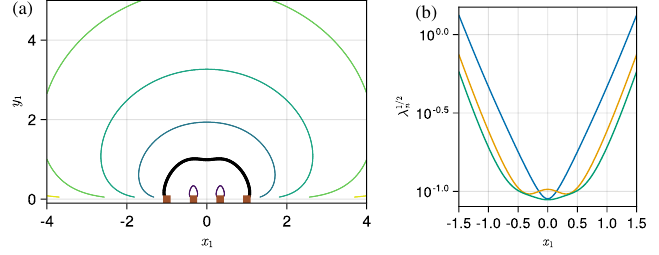}
\end{center}
\caption{One true vortex and one-vortex estimator, with noise $\noisestddev = 5\times 10^{-4}$. (a) Contours (on a log scale) of maximum length of covariance ellipsoid, $\eigenSx_{\statedim}^{1/2}$, as a function of true vortex position $(x_{1},y_{1})$ using four sensors, shown as brown squares. Contours are $10^{-2}$ through $10^{2}$, with $10^{-1}$ depicted with a thick black line. (b) Maximum length of covariance ellipsoid versus horizontal position of true vortex (with vertical position held fixed at $y_{1} = 1$), and varying number of sensors along line segment $x = [-1,1]$ ($d=3$, blue; $d = 4$, gold; $d = 5$, green)} \label{fig:maxcovar-vort-1vortex}
\end{figure}

\subsubsection{Effect of the true vortex position}

It is particularly important to explore how the uncertainty is affected by the position of the true vortex relative to the sensors. We address this question here by varying this position relative to a fixed set of sensors with a fixed level of noise, $\noisestddev = 5\times 10^{-4}$. Figure~\ref{fig:maxcovar-vort-1vortex}(a) depicts contours (on a log scale) of the resulting maximum length of the covariance ellipsoid, $\eigenSx_{\statedim}^{1/2}$, based on four sensors placed on the $x$ axis between $-1$ and $1$. The contours reveal that there is little uncertainty when the true vortex is in the vicinity of the sensors, but the uncertainty increases sharply with distance when the vortex lies outside the extent of sensors. Indeed, one finds empirically that the rates of increase scale approximately as $\eigenSx_{\statedim}^{1/2} \sim |y_{1}|^{5}$ and $\eigenSx_{\statedim}^{1/2} \sim |x_{1}|^{6}$. This behavior does not change markedly if we vary the number of sensors, as illustrated in Figure~\ref{fig:maxcovar-vort-1vortex}(b). As the true vortex's $x$ position varies (and $y_{1}$ is held constant at 1), there is a similarly sharp rate of increase outside of the region near the sensors for 3, 4, or 5 sensors. However, though there is a small range of positions near $x_{1} = 0$ in which 3 sensors have less uncertainty than 4, there is generally less uncertainty at all vortex positions with increasing numbers of sensors. Furthermore, the uncertainty is less variable in this near region when 4 or 5 sensors are used.


\begin{figure}
\begin{center}
\includegraphics{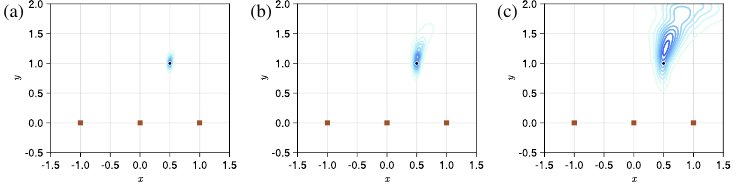}
\end{center}
\caption{One true vortex and one-vortex estimator, using three sensors (shown as brown squares), with noise levels (a) $\noisestddev = 2.5 \times 10^{-4}$, (b) $5\times 10^{-4}$, and (c) $1\times 10^{-3}$. Each panel depicts contours of expected vorticity field. True vortex position shown as filled black circle in each.} \label{fig:noise-vort-1vortex}
\end{figure}

\subsubsection{Effect of sensor noise}

From the derivation in \sect~\ref{sec:gauss1}, we already know that the true covariance should depend linearly on the noise variance \eqref{eq:eigenSx}. In this section, we explore the effect of sensor noise on the estimation of a single vortex using MCMC and the subsequent fitting with a Gaussian mixture model. We keep the number of sensors fixed at 3 arranged along a line between $x = -1$ and $1$, and the true vortex in the original configuration, $(x_{1},y_{1},\Gamma_{1}) = (0.5,1,1)$. Figure~\ref{fig:noise-vort-1vortex} depicts the expected vorticity field as the noise standard deviation $\noisestddev$ increases. Unsurprisingly, the expected vorticity distribution exhibits increasing breadth as the noise level increases. However, it is notable that this breadth becomes increasingly directed away from the sensors as the noise increases. Furthermore, the center of the distribution lies somewhat further from the sensors than the true state, indicating a bias error. This trend toward increased bias error with increasing sensor noise is also apparent in other sensor numbers and arrangements.



\subsubsection{Effect of true vortex radius}

\begin{figure}
\begin{center}
\includegraphics[width=\textwidth]{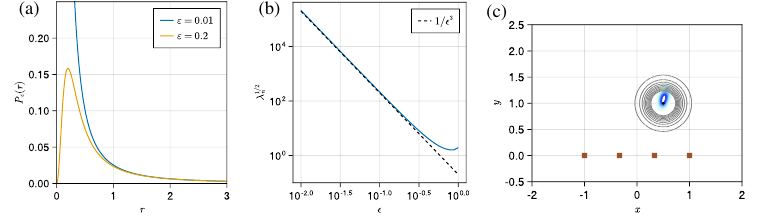}
\end{center}
\caption{(a) Regularized vortex-pressure kernel $\vorkernbase_{\blobrad}$, with two different choices of blob radius. (b) The maximum length of uncertainty ellipsoid versus blob radius $\blobrad$, for one true vortex and four sensors, when blob radius is included as part of the state vector, for a true vortex at $(x_{1},y_{1},\Gamma_{1}) = (0.5,1,1)$. (c) Vorticity contours for a true vortex with radius $\blobrad=0.2$ (in gray) and expected vorticity from a vortex estimator with radius $\blobrad = 0.01$ (in blue), using 4 sensors (shown as brown squares). } \label{fig:blobradius}
\end{figure}

Throughout most of this paper, the radius of the true vortices is fixed at $\blobrad = 0.01$, and the estimator vortices share this same radius. However, for practical application purposes, it is important to explore the extent to which the estimation is affected by a mismatch between these. If the true vortex is more widely distributed, can an estimator consisting of a small-radius vortex reliably determine its position and strength? Furthermore, can the radius itself be inferred? These two questions are closely related, as we will show. First, it is useful to illustrate the effect of the vortex radius on the pressure field associated with a vortex, as in Figure~\ref{fig:blobradius}(a), which shows the vortex-pressure kernel $\vorkernbase_{\blobrad}$ for two different vortex radii, $\blobrad = 0.01$ and $\blobrad = 0.2$. As this plot shows, the effect of vortex radius is fairly negligible beyond a distance of 5 times the larger of these two vortex radii.

As a result of this diminishing effect of vortex radius, one expects that it is very challenging to estimate this radius from pressure sensor measurements outside of the vortex core. Indeed, that is the case, as Figure~\ref{fig:blobradius}(b) shows. This figure depicts the maximum length of the covariance ellipsoid as a function of true vortex radius, when four sensors along the $x$ axis are used to estimate this radius (in addition to vortex position and strength), for a true vortex at $(0.5,1)$ with strength 1. The uncertainty is far too large for the radius to be observable until this radius approaches $\blobrad = 1$. Even when the sensors are within the core of the vortex, they are confused between the blob radius and other vortex states. In fact, one can show that the dependence of the maximum uncertainty on $\blobrad^{-3}$ arises because of the nearly identical sensitivity that the pressure sensors have to changes of blob radius and changes in other states, e.g., vertical position in this case. The leading order term of the difference is proportional to $\blobrad^{3}$. Of course, if we presume precise knowledge of the other states, then vortex radius becomes more observable.

The insensitivity of pressure to vortex radius has a very important benefit, because it ensures that, even when the true vortex is relatively broad in size, a vortex estimator with small radius can still reliably infer the vortex's position and strength. This is illustrated in Figure~\ref{fig:blobradius}(c), which depicts the contours of a true vortex of radius $\blobrad = 0.2$ (with the same position and strength as in panel (b)), and the expected vorticity contours from an estimate carried out with a vortex element of radius $\blobrad = 0.01$. It is apparent from this figure that the center of the vortex is estimated very well. In fact, the mean of the MCMC samples is $(\overline{x}_{1},\overline{y}_{1},\overline{\Gamma}_{1}) = (0.51, 1.08, 1.04)$, quite close to the true values.

\subsection{Inference of multiple vortices}

The previous sections have illustrated the crucial aspects of estimating a single vortex. In this section, we focus on inferring multiple vortices. As in the case with a single vortex, the eigenvalues and eigenvectors of the true covariance ellipsoid $\approxstatecovar$ will serve an important role in revealing many of the challenges in this context. However, some of the multiple vortex cases will have multiple possible solutions, and we must rely on the MCMC samples to reveal these. In the examples carried out in this section, we use the same uniform prior as in the previous section, except that now the prior vortex strengths can be negative (lying between $(-2,2)$).

\begin{figure}
\begin{center}
\includegraphics[width=\textwidth]{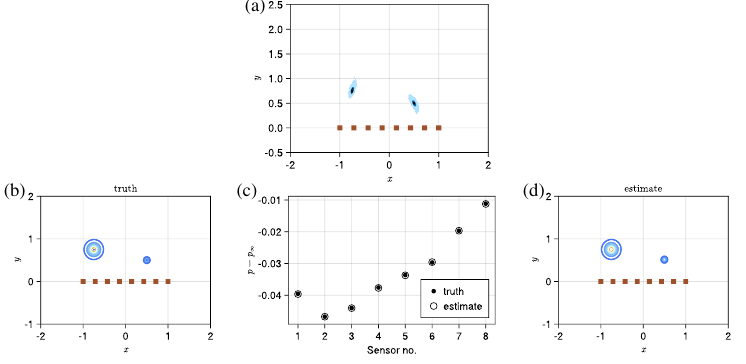}
\end{center}
\caption{True vortex configuration $(x_{1},y_{1},\Gamma_{1}) = (-0.75,0.75,1.2)$ and $(x_{2},y_{2},\Gamma_{2}) = (0.5,0.5,0.4)$, using two-vortex estimator with eight sensors. (a) MCMC samples. True vortex positions shown as filled black circle in each. Ellipses correspond to the estimated covariance of the samples. (b) True pressure field, (c) the comparisons of sensor values between truth (filled circles) and estimate (open circles), and (d) estimated pressure field. } \label{fig:2vortex}
\end{figure}

\subsubsection{Two true vortices with a two-vortex estimator}

In this section, we will study the inference of a pair of vortices using a two-vortex estimator. The basic configuration of true vortices consists of $(x_{1},y_{1},\Gamma_{1}) = (-0.75,0.75,1.2)$ and $(x_{2},y_{2},\Gamma_{2}) = (0.5,0.5,0.4)$, both of radius $\blobrad = 0.01$. As in the previous section, sensors have noise $\noisestddev = 5\times 10^{-4}$. In Figure~\ref{fig:2vortex} we demonstrate the estimation of this pair of vortices with eight sensors. The estimator is very effective at inferring the locations and strengths of the individual vortices: the mean state of the samples is $(\overline{x}_{1},\overline{y}_{1},\overline{\Gamma}_{1}) = (-0.75,0.77,1.23)$ and $(\overline{x}_{2},\overline{y}_{2},\overline{\Gamma}_{2}) = (0.50,0.50,0.39)$, and the pressure field predicted by the expected state matches well with the true pressure field.

\begin{figure}
\begin{center}
\includegraphics[width=0.9\textwidth]{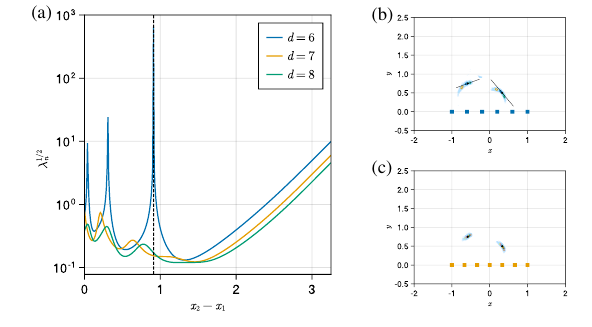}
\end{center}
\caption{Two true vortices with strengths $\Gamma_{1} = 1.2$ and $\Gamma_{2} = 0.4$ and two-vortex estimator. (a) Maximum length of uncertainty ellipsoid for varying horizontal distance between true vortices and varying number of sensors. (b,c) The MCMC samples from a two-vortex estimator, for the true vortex configuration with peak uncertainty at $x_{2} - x_{1} = 0.91$ (shown as a dashed vertical line in (a)), using (b) six sensors and (c) seven sensors. True vortex positions are shown as filled black circles in each. The thin lines in the six-sensor case indicate the direction of maximum uncertainty for each vortex position.} \label{fig:maxcovar-2vortex}
\end{figure}

In this basic configuration, the vortices are widely separated so that the estimator's challenge is similar to that of two isolated single vortices, each estimated with four sensors. However, unique challenges arise as the true vortices become closer, as Figure~\ref{fig:maxcovar-2vortex} shows. Here, we keep the strength and vertical position of each true vortex the same as in the basic case, but vary both vortices' horizontal position---the left one is moved rightward and the right one is moved leftward---in such a manner that their average is invariant, $(x_{1}+x_{2})/2 = -0.125$. Three different numbers of sensors are used, $\measdim = 6, 7, 8$, all uniformly distributed between $[-1,1]$. In Figure~\ref{fig:maxcovar-2vortex}(a), it is clear that using six sensors, though ostensibly sufficient to estimate the six states, is actually insufficient in a few isolated cases in which the maximum uncertainty becomes infinite. These cases are examples of rank deficiency in the vortex estimator. Importantly, this rank deficiency disappears when more than six sensors are used. An example of the estimator's behavior in one of these rank-deficient configurations is depicted in Figure~\ref{fig:maxcovar-2vortex}(b,c). When six sensors are used (panel (b)), the MCMC samples are distributed more widely, along a manifold in the vicinity of the true state, with the eigenvector of the most-uncertain eigenvalue tangent to this manifold. However, when seven sensors are used (panel (c)), the MCMC samples are more tightly distributed around the true state.

\begin{figure}
\begin{center}
\includegraphics[width=0.8\textwidth]{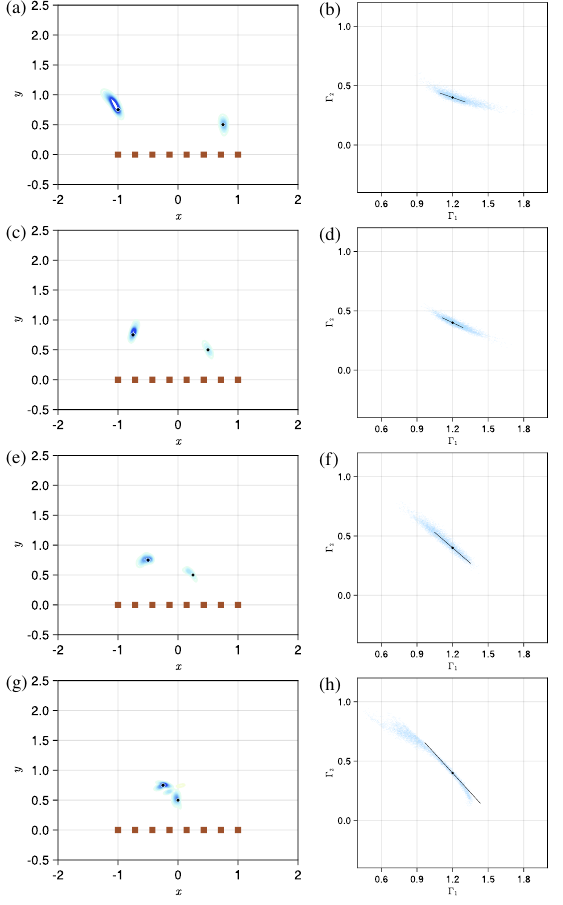}
\end{center}
\caption{Two true vortices with strengths $\Gamma_{1} = 1.2$ and $\Gamma_{2} = 0.4$, and two-vortex estimator with 8 sensors (shown as brown squares in left panels). Varying separation between vortices: (a,b) $x_{2} - x_{1} = 1.75$, (c,d) $1.25$, (e,f) $0.75$, (g,h) $0.25$. Each left panel depicts contours of expected vorticity field, with true vortex positions shown as filled black circle in each. Each right panel depicts MCMC samples of vortex strengths, with true vortex strengths shown as black circle, and the longest axis of the true covariance ellipsoid depicted by the line.} \label{fig:varying-sep-2vortex}
\end{figure}

Thus, we can avoid rank deficiency by using more sensors than states. As a demonstration, we show in the left panels of Figure~\ref{fig:varying-sep-2vortex} the expected vorticity field that results from estimating four different true vortex configurations with eight sensors. In each case, the locations of the vortices are accurately estimated with relatively little uncertainty, even as the vortices become closer to each other than they are to the array of sensors. However, with closer vortices there is considerable uncertainty in estimating the strengths of the individual vortices, as exhibited in the right panels of Figure~\ref{fig:varying-sep-2vortex}, each corresponding to the vortex configuration on the left.

\begin{figure}
\begin{center}
\includegraphics[width=0.8\textwidth]{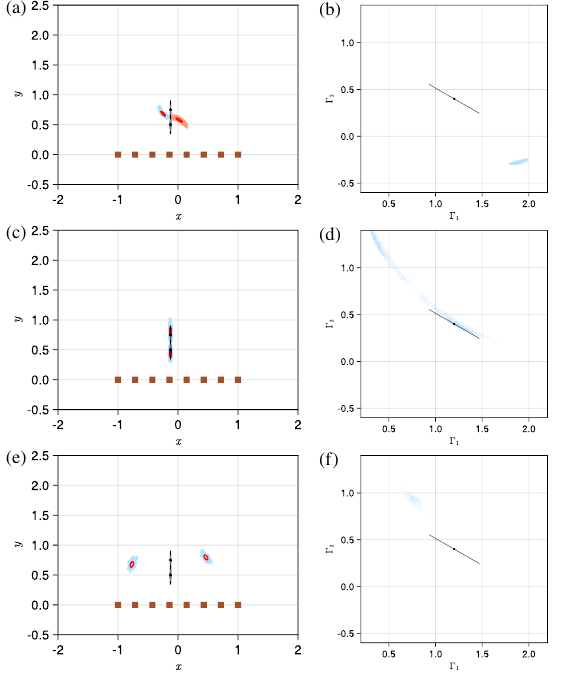}
\end{center}
\caption{Two true vortices $(x_{1},y_{1},\Gamma_{1}) = (-0.125,0.75,1.2)$ and $(x_{2},y_{2},\Gamma_{2}) = (-0.125,0.5,0.4)$, and a two-vortex estimator with 8 sensors (shown as brown squares). Each row depicts one mixture model component. In each left panel, true vortex positions are shown as filled black circles, and each vortex's corresponding position covariance is shown as filled gray ellipse. Red ellipses depict the position covariance of the mixture model component and the dots are the MCMC samples with greater than 50 percent probability of belonging to that component (blue for positive strength; red for negative). Right panels depict the MCMC samples of vortex strengths, with true vortex strengths shown as black circle in each, and the longest axis of the true covariance ellipsoid depicted by the line.} \label{fig:closest-sep-2vortex}
\end{figure}

As the true vortices become even closer than in the examples in Figure~\ref{fig:varying-sep-2vortex}, multiple solutions emerge. This is illustrated in Figure~\ref{fig:closest-sep-2vortex}, depicting the extreme case of one vortex just above the other. The MCMC identifies three modes of the posterior, each representing a different candidate solution for the estimator. One mode consists of vortices of opposite sign to either side of the true set, shown in the top row of Figure~\ref{fig:closest-sep-2vortex}. The second mode, in the middle row, comprises vortices very near the true set, though the strengths of the vortices are quite uncertain, as evidenced by the long ridge of samples in the strength plot in Figure~\ref{fig:closest-sep-2vortex}(d). Finally, the bottom row shows a mode that has positive vortices further apart than in the other two modes.

It is natural to ask whether we can prefer one of these two candidate solutions over the other. One way to do so is to assess them based on their corresponding weights $\gmmwt_{k}$ in the mixture model, since each of these represents the probability of a given sample point belonging to that component. However, interpreting the mixture model's weights in this fashion requires that the MCMC has reached equilibrium, which can be challenging to determine with multimodal sampling. Instead, we follow the intuition that, if a mode is to be considered a more likely solution of the inference problem than another mode, then the samples belonging to that mode should be closer to the true observation. For this assessment we can compare the maxima of the log-posterior (\ref{eq:logprob}) among the samples belonging to two modes. The mode with a significantly larger maximum (i.e., significantly closer to zero, since (\ref{eq:logprob}) is non-positive) is a superior candidate solution. For the two modes shown in Figure~\ref{fig:closest-sep-2vortex}, the maximum log-posteriors are $-0.20$, $-0.11$, and $-14.83$, respectively, suggesting that the mode in the middle row is mildly superior to that of the top row and clearly superior to that of the bottom row. Indeed, the fact that this clearly inferior mode appears among the samples at all is likely due to incomplete MCMC sampling. Thus, in this example with two very closely spaced positive-strength vortices, the true solution is discernible from the two spurious solutions.

\begin{figure}
\begin{center}
\includegraphics[width=0.8\textwidth]{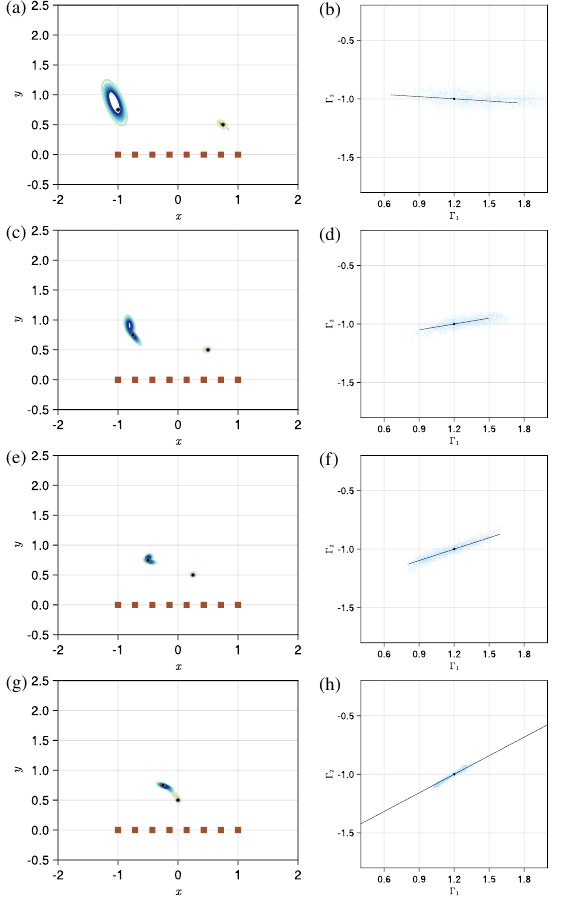}
\end{center}
\caption{Two true vortices with strengths $\Gamma_{1} = 1.2$ and $\Gamma_{2} = -1.0$, and two-vortex estimator with 8 sensors  (shown as brown squares in left panels). Varying separation between vortices: (a,b) $x_{2} - x_{1} = 1.75$, (c,d) $1.25$, (e,f) $0.75$, (g,h) $0.25$. Each left panel depicts contours of expected vorticity field, with true vortex positions shown as filled black circle in each. Each right panel depicts MCMC samples of vortex strengths, with true vortex strengths shown as black circle, and the longest axis of the true covariance ellipsoid depicted by the line.} \label{fig:varying-sep-2vortex-GpGm}
\end{figure}

\begin{figure}
\begin{center}
\includegraphics[width=0.8\textwidth]{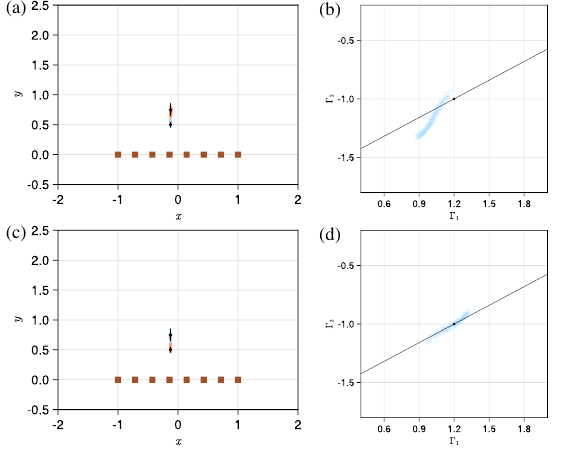}
\end{center}
\caption{Two true vortices $(x_{1},y_{1},\Gamma_{1}) = (-0.125,0.75,1.2)$ and $(x_{2},y_{2},\Gamma_{2}) = (-0.125,0.5,-1.0)$, and a two-vortex estimator with 8 sensors (shown as brown squares). Each row depicts one mixture model component. In each left panel, true vortex positions are shown as filled black circles, and each vortex's corresponding position covariance is shown as filled gray ellipse. Red ellipses depict the position covariance of the mixture model component and the blue dots are the MCMC samples with greater than 50 percent probability of belonging to that component (blue for positive strength; red for negative). Right panels depict the MCMC samples of vortex strengths, with true vortex strengths shown as black circle in each, and the longest axis of the true covariance ellipsoid depicted by the line.} \label{fig:closest-sep-2vortex-GpGm}
\end{figure}

In Figure~\ref{fig:varying-sep-2vortex-GpGm} we carry out the same procedure of bringing two vortices closer together as in Figure~\ref{fig:varying-sep-2vortex}, but now we do so for one vortex of positive strength ($1.2$) and another of negative strength ($-1.0$). We get similar results as before, successfully estimating the vortex locations and strengths. The most challenging case among these is the first, in which the two vortices are furthest apart and near the extreme range of the sensors. Interestingly, no spurious solutions arise as the vortices become very close together, as they did in the previous example. In fact, when the two opposite-sign vortices are vertically aligned, as in Figure~\ref{fig:closest-sep-2vortex-GpGm}, the estimator has no difficulty in identifying the individual vortices and their strengths. The figure ostensibly depicts two modes identified by the estimator, but in fact these modes are identical aside from the sign of their strengths. They remain distinct to the estimator only because they have eluded our simple mitigations for the relabeling and strength symmetries (of ordering the vortices by their $x$ position and assuming that the leftmost has positive strength). Clearly we could have chosen a different mitigation strategy to avoid this, but we include the separate modes here for illustration purposes.  

\subsubsection{Three true vortices}

\begin{figure}
\begin{center}
\includegraphics[width=\textwidth]{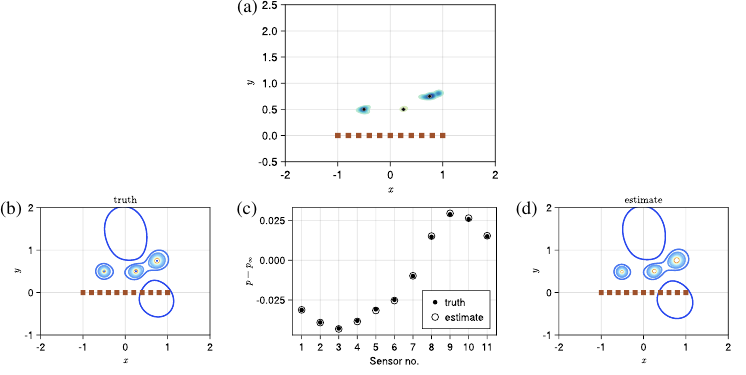}
\end{center}
\caption{Three true vortices with states $\state_{1} = (x_{1},y_{1},\Gamma_{1}) = (-0.5,0.5,1)$, $\state_{2} = (0.25,0.5,-1.2)$ and $\state_{3} = (0.75,0.75,1.4)$, respectively, and a three-vortex estimator using eleven sensors, shown as brown squares. (a)) Expected vorticity field, (b) true pressure field, (c) the comparisons of sensor values between truth (filled circles) and estimate (open circles), and (d) estimated pressure field.}\label{fig:3vortex-vorticity}
\end{figure}

In this section we demonstrate the performance of the estimator on cases with three true vortices. In the first example, we will use three vortices in the estimator. Let the true state comprise $\state_{1} = (x_{1},y_{1},\Gamma_{1}) = (-0.5,0.5,1)$, $\state_{2} = (0.25,0.5,-1.2)$ and $\state_{3} = (0.75,0.75,1.4)$. Here, we apply the techniques shown to enhance performance in the previous sections: we use eleven sensors, two more than the marginal number, to avoid rank deficiency; and we choose a top candidate among the identified modes based on the maximum value of the log-posterior. The resulting estimated solution has only a single candidate mode, whose mean is $\meanstate_{1} = (-0.47,0.53,1.10)$, $\meanstate_{2} = (0.25,0.51,-1.25)$, $\meanstate_{3} = (0.68,0.75,1.36)$. This is shown in Figure~\ref{fig:3vortex-vorticity}. Both the expected vorticity field and the pressure field are captured very well by the estimator.

\begin{figure}
\begin{center}
\includegraphics[width=0.8\textwidth]{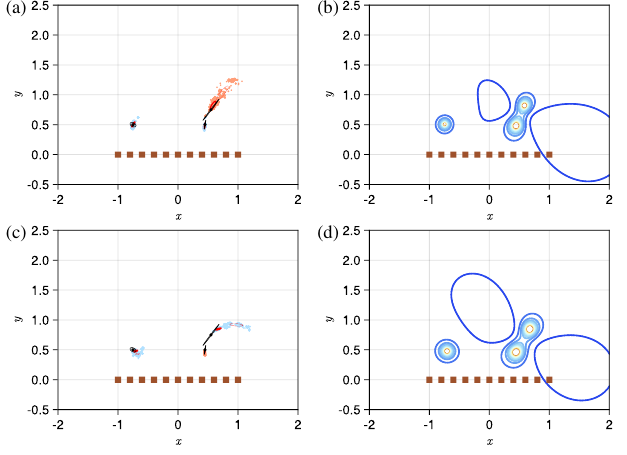}
\end{center}
\caption{Three true vortices with states $\state_{1} = (x_{1},y_{1},\Gamma_{1}) = (-0.75,0.5,1)$, $\state_{2} = (0.45,0.5,-1.2)$ and $\state_{3} = (0.55,0.75,1.4)$, and a three-vortex estimator with 11 sensors (shown as brown squares). Each row depicts a mixture model component. In each left panel, true vortex positions are shown as filled black circles, and each vortex's corresponding position covariance is shown as filled gray ellipse. Red ellipses depict the position covariance of the mixture model component and the blue dots are the MCMC samples with greater than 50 percent probability of belonging to that component (blue for positive strength; red for negative). Right panels depict contours of the estimated pressure field for that mode.} \label{fig:closest-sep-3vortex}
\end{figure}

It is important to note that, in the previous example, the estimator identifies another mode (not shown) in which the signs of the rightmost two vortices are switched. However, this candidate solution was discarded on the basis of the maximum log-posterior criterion: the selected mode's value is $-0.72$, while the discarded mode's is smaller, $-0.85$. This slight difference is entirely due to the non-linear coupling between the vortices via the interaction kernel. By restricting the leftmost vortex to be positive, the signs of the middle and rightmost vortices are established through this coupling. Because the algebraic decay of both types of kernels in equation \eqref{eq:pvpfull0} is the same, the ability to discriminate the signs of the vortices requires a degree of balance between the vortex positions and the sensors. As a counterexample, when two of the three vortices form a compact pair that is well-separated from the third, the estimator tends to be less able to prefer one choice of sign for the vortex pair over the other. An example is shown in Figure~\ref{fig:closest-sep-3vortex}, in which the rightmost pair of vortices has opposite sign in each mode. The corresponding pressure fields shown on the right are nearly identical because the coupling of the pair with the leftmost vortex is much weaker than in the pair itself. Thus, these modes are indistinguishable by our maximum log-posterior criteria: in the absence of additional prior knowledge, we cannot discern one from the other.

\begin{figure}
\begin{center}
\includegraphics[width=\textwidth]{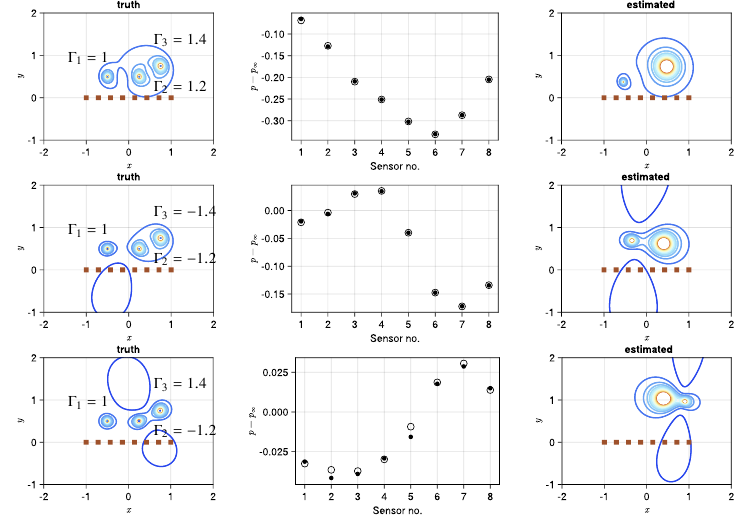}
\end{center}
\caption{Estimated pressure fields for three true vortices with positions $(x_{1},y_{1}) = (-0.5,0.5)$, $(0.25,0.5)$ and $(0.75,0.75)$, using a two-vortex estimator with 8 sensors (shown as brown squares). (Top row) True vortex strengths $\Gamma_{1} = 1$, $\Gamma_{2} = 1.2$, and $\Gamma_{3} = 1.4$. (Center row) $\Gamma_{1} = 1$, $\Gamma_{2} = -1.2$, and $\Gamma_{3} = -1.4$, (Bottom row) $\Gamma_{1} = 1$, $\Gamma_{2} = -1.2$, and $\Gamma_{3} = 1.4$. Rightmost panels depict contours of the estimated pressure field for each case.} \label{fig:2vortex-3truth}
\end{figure}

The three-vortex estimator must explore a 9-dimensional space for the solution, a challenging task even with the various MCMC and symmetry mitigation techniques we have used in this paper. Thus, it is useful to restrict the estimator to search a lower-dimensional space, and the easiest way to achieve this is by using fewer vortices in the estimator. In Figure~\ref{fig:2vortex-3truth}, we illustrate the behavior a two-vortex estimator on the three-vortex configuration in Figure~\ref{fig:3vortex-vorticity}, in variations in which the signs of the right two true vortices are changed. The range of strengths in the prior is expanded in this problem to $(-4,4)$. In the first case, the true vortex strengths are all positive. The two-vortex estimator identifies a single mode, with mean vortex states $\meanstate_{1} = (-0.54, 0.37, 0.64)$ and $\meanstate_{2} = (0.48, 0.74, 3.25)$, and a maximum log-posterior of $-30.7$. In other words, the estimator places one vortex near (and slightly weaker than) the leftmost true vortex, and another vortex near the center of the rightmost pair, with a strength roughly equal to the sum of the pair. In the second case, the two rightmost vortices are both negative, and the estimator produces an analogous result, aggregating the two negative vortices into a single vortex. The estimated state is $\meanstate_{1} = (-0.34, 0.69, 1.27)$ and $\meanstate_{2} = (0.40, 0.62, -3.07)$, with maximum log-posterior $-20.4$, so that the rightmost pair is once again approximated by a single vortex with roughly the sum of the pair's strength.

The third case is the most interesting. Here, the true vortex configuration consists of positive, negative, and positive vortices from left to right, so there is no pairing of like-sign vortices as in the previous two cases. The estimator identifies a solution consisting of $\meanstate_{1} = (0.40, 1.03, 3.50)$ and $\meanstate_{2} = (0.90, 0.97, -1.2)$. Neither of these vortices bears an obvious connection with one of the true vortices, so no aggregation is possible. The estimator has done the best in can in the lower-dimensional space available to it, aliasing the true flow onto a dissimilar flow state. The maximum log-posterior is $-88.6$, significantly lower than in the other two cases.

\section{Conclusions}
\label{sec:conclude}

In this paper, we have explored the inference of regularized point vortices from a finite number of pressure sensors with noisy measurements. By expressing the problem in a Bayesian (probabilistic) manner, we have been able to quantify the uncertainty of the estimated vortex state and to explore the multiplicity of possible solutions, which are expressed as multiple modes in the posterior distribution.  We sampled the posterior with Markov-chain Monte Carlo and applied Gaussian mixture modeling to develop a tractable approximation for the posterior from the samples. Mixture modeling allowed us to soft-classify the samples into each mode. We reduced the multiplicity by anticipating many of the symmetries that arise in this inference problem---strength, relative position, and vortex re-labeling---and then mitigated their influence through simple techniques: e.g., restricting the prior region, strictly ordering the vortices in the state vector by $x$ coordinate. The remaining multiplicity of solutions were identified by thoroughly exploring the prior region with help from the method of parallel tempering in MCMC. Where possible, the best candidate solution was discerned by monitoring the maximum log-posterior in each mode. We have also made use of the largest eigenvalue and associated eigenvector of the true covariance matrix in order to illuminate many of the challenges of the inference.

On a variety of configurations of one, two, or three true vortices, we have made several observations of this vortex inference problem. One must use at least as many sensors as there are estimator states in order to infer a unique vortex system rather than a manifold of equally-possible states. Using one additional sensor guards against the risk of cases of rank deficiency, which arise occasionally when multiple vortices are used in the estimator. However, additional sensors do not significantly improve the uncertainty of the estimate. Uncertainty scales linearly with sensor noise. It also rises very rapidly, with the fifth or sixth power of distance, when the true vortex lies outside of a region of the sensors. The size of the vortex is exceptionally challenging to estimate because its effect on pressure is almost indistinguishable from other vortex states. However, this fact is also advantageous, for it allows us to use a small radius (nearly-singular) vortex to accurately estimate the position and strength of a larger one. For systems of multiple vortices, the estimator relies on the non-linear coupling between them to ascertain the sign of the strength of each. Even when multiple modes emerge, one can often discern the best candidate among the modes based on the criterion of maximum probability (i.e., the shortest distance to the true measurements). This approach fails in some cases when the vortices are imbalanced, such as when a pair of vortices is well separated from a third. When the estimator uses fewer vortices than in the true configuration, it identifies the most likely solution in the reduced state space. Often, this reduced-order estimate appears to be a natural aggregation of the true vortex state, but in some cases the estimator aliases the sensors onto a dissimilar vortex configuration when no aggregated one is possible.

It is important to reiterate that the static inference we have studied in this paper \change{is useful both in its own right as a one-time estimate of the vorticity field and as part of a sequential estimation of a time-varying flow. In the latter context---for example, in an ensemble Kalman filter 
\citep{Silva2018,darakjde2018,provost2021ensemble,leprovost2022lrenkf}---the inference comprises the analysis part of every step, when data is assimilated into the prediction. Some of the challenges and uncertainty of the static inference identified in this paper are overcome with advancing time as the sensors observe the evolving configuration of the flow and the forecast model predicts the state's evolution. Indeed, we speculate that the rank deficiency is partially mitigated in this manner. However, as new vortices are generated or enter the region of interest, the prior distribution obtained from the previous step will not be descriptive of these new features. Thus, our assumption of a non-informative prior remains relevant ever after the initial step of a flow, and the conclusions we have drawn in this work can guide an estimator that remains receptive to new vortex structures. Overall, most of the conclusions of this paper, including rank deficiency, the decay of signal strength with distance, and the effects of vortex couplings, are impactful on a sequential filter's overall performance.  It is also important to stress the fact that, when the vortex state can be unambiguously inferred from a set of sensors, it implies that the sensor data contain sufficient information to describe the flow. This has implications for reinforcement learning-based flow control approaches \citep{Verma2018}, which treat the fluid flow as a partially observable Markov decision process. Pressure sensor data potentially make the process more Markovian, and therefore more amenable to learning a control policy \citep{Renn2022}. In short, there is less risk that control decisions made from measured pressures data are working on a mistaken belief about the flow state.}


\change{Finally, though we have addressed many substantive questions in this work, there are still several to address: In a realistic high Reynolds number flow}, i.e. comprising a few dominant coherent structures amidst shear layers and small-scale vortices, can the estimator infer the dominant vortices? \change{Our work here has exhibited some key findings that suggest that it can; in particular, weaker vortices with comparatively little effect on the velocity and small measured pressures are neglected by the estimator in favor of stronger vortices. But this question deserves more thorough treatment. Also, how does the presence of a body affect the vortex estimation?  Furthermore, when such a body is in motion, or subject to a free stream, can the vortices and the body motion be individually inferred? In the presence of a body, stationary or in motion, it is straightforward to expand the pressure--vortex relation presented here to develop an inviscid observation model that incorporates geometry, other flow contributors, and their couplings (e.g., between a vortex and a moving body). The estimation framework presented here can readily accommodate such an enriched model. Indeed, in our prior work with an ensemble Kalman filter \citep{darakjde2018,provost2021ensemble}, we have found that the signal of a vortex can be enhanced in pressure sensors on a nearby wall due to the effect of the vortex's image, and that this inviscid observation model remains effective in a viscous setting, such as flow past a flat plate. However, there are challenges with estimating vortices near regions of large wall curvature, such as the edges of an airfoil, where large pressure gradients, viscous effects, and the subsequent generation of new vortices complicate the flow process. In our previous work, we have approximately accounted for these processes by augmenting the state with a leading-edge suction parameter \citep{ramesh14}. However, it would also be worthwhile to investigate approaches in which the physics-based observation model is replaced or augmented with a data-driven approach, such as a trained neural network \citep{Zhong2023}.}

\section*{Acknowledgements}
Support for this work by the National Science Foundation under Award number 2247005 is gratefully acknowledged.

\section*{Declaration of Interests}
The authors report no conflict of interest.


\appendix

\section{Pressure and vorticity in unbounded flow}
\label{sec:unbound}

\subsection{Pressure from vorticity in unbounded flow}

We start by writing the Poisson equation for pressure, obtained by taking the divergence of the incompressible Navier--Stokes equations and using the fact that velocity is divergence-free. To emphasize the role of vorticity, we first re-write the convective term in the equations with the help of the vector identity $\v\cdot \nabla\v = \nabla |\v|^{2}/2 - \v \times \w$, where $\w = \nabla\times \v$, obtaining
\begin{equation}
\lap \left(p + \frac{1}{2} \rho |\v|^{2}\right) = \rho \nabla\cdot \left(\v \times \w\right).
\label{eq:ppevort}
\end{equation}
The quantity $\v\times \w$ is the so-called Lamb vector \citep{lamb:1j}. To formally solve this problem for the quantity in parentheses, we can make use of the free-space Green's function for the (negative) Laplace operator, i.e., the solution of
\begin{equation}
\lap \green = -\delta(\rvec),
\end{equation}
so that, by Green's theorem (and integration by parts), we get the pressure to within a uniform constant, $C$\footnote{Actually, any homogeneous solution of Laplace's equation could be added to this expression, but in this unbounded domain only a uniform constant will permit a finite pressure at infinity.}:
\begin{equation}
\press(\rvec) = C -\frac{1}{2} \rho |\v(\rvec)|^{2} - \rho \int \left(\v(\rvec') \times \w(\rvec')\right)\cdot \nabla\green(\rvec - \rvec') \,\dvol(\rvec'),
\label{eq:press1}
\end{equation}
where $\dvol(\rvec')$ denotes the infinitesimal volume element at point $\rvec'$, and the integral is taken over the entire space. Thus, the pressure at any observation point $\rvec$ is partly attributable to the velocity magnitude at $\rvec$ (the so-called dynamic pressure) and partly induced by the Lamb vector field. The dynamic pressure's role is simple and familiar: faster local flow speed is associated with lower pressure. The role of the Lamb vector is less familiar, so it is helpful to manipulate this integral further.

The Lamb vector field is clearly only distributed over regions that are vortical. To further elucidate the role of vorticity, we will make use of the fact that the velocity can itself be recovered from the vorticity field, via the Biot--Savart integral, plus any additional irrotational contributions represented by a scalar potential field. In an unbounded context this irrotational contribution is a possible uniform flow, $\Vinf$. Thus,
\begin{equation}
\v(\rvec) = \Vinf + \vw(\rvec),
\label{eq:vhelm}
\end{equation}
where
\begin{equation}
\vw(\rvec) \equiv \int \nabla \green(\rvec - \rvec') \times \w(\rvec')\,\dvol(\rvec').
\label{eq:vbiot}
\end{equation}

For clarity in what follows, it is useful to denote the contribution from each infinitesimal volume element by
\begin{equation}
\dvw{\rvec}{\rvec'} \equiv \nabla \green(\rvec - \rvec') \times \w(\rvec')\,\dvol(\rvec'),
\label{eq:dvw}
\end{equation}
so the Biot-Savart integral can be written more compactly as
\begin{equation}
\vw(\rvec) = \int_{\rvec'} \dvw{\rvec}{\rvec'},
\label{eq:vbiot2}
\end{equation}
where the notation indicates that we are taking this integral over all points $\rvec'$ in space. Interestingly, this infinitesimal velocity contribution also appears in the Lamb vector integral in \eqref{eq:press1}, made apparent by permuting the triple product in the integral so that we can write it alternatively as
\begin{equation}
\press(\rvec) = C -\frac{1}{2} \rho |\v(\rvec)|^{2} + \rho \int_{\rvec'} \v(\rvec') \cdot \dvw{\rvec}{\rvec'}.
\label{eq:press1-1}
\end{equation}
To eliminate the constant $C$, we note that, provided the uniform flow is steady, the limits as $|\rvec| \rightarrow \infty$ are $\press(\rvec) \rightarrow \press_{\infty}$ (the ambient pressure), $\v(\rvec) \rightarrow \Vinf$, and $\dvw{\rvec}{\rvec'} \rightarrow 0$. Thus, $C = \press_{\infty} + \frac{1}{2} \rho |\Vinf|^{2}$, and
\begin{equation}
\press(\rvec) - \press_{\infty} = \frac{1}{2} \rho \left(|\Vinf|^{2} - |\v(\rvec)|^{2}\right) + \rho \int_{\rvec'} \v(\rvec') \cdot \dvw{\rvec}{\rvec'}.
\label{eq:press1-2}
\end{equation}
This form of the pressure is reminiscent of the familiar Bernoulli equation. In fact, in the special case of an inviscid flow comprising singular distributions of vorticity (e.g., point vortices or vortex filaments), then it can be shown that the integral is equivalent to $-\rho \partial \spot/\partial t$, where $\spot$ is the equivalent scalar potential field induced by the singular vortices. The unsteadiness of this potential field at the observation point $\rvec$ is brought about by the convection and tilting of the vortices by their local velocity field, associated with the presence of $\v$ inside the integral. Thus, under those special circumstances, \eqref{eq:press1-2} is identically the Bernoulli equation, as one would expect. For the general case of distributed vorticity, the equation no longer reverts to Bernoulli, but the pressure still receives an essential `velocity-modulated' contribution from the vorticity.

\begin{figure}
\begin{center}
\includegraphics{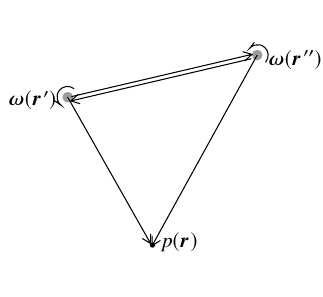}
\caption{Triadic interaction between two vorticity-laden elements and the pressure at some point.} \label{fig:triadic}
\end{center}
\end{figure}

We could finish our derivation with expression \eqref{eq:press1-2}. However, in order to distinguish the contributions from the uniform flow and the vorticity, we introduce \eqref{eq:vhelm} for the velocity. Simplifying the resulting expression, we arrive at
\begin{equation}
\press(\rvec) - \press_{\infty} = -\frac{1}{2} \rho |\vw(\rvec)|^{2} + \rho \int_{\rvec'} \vw(\rvec') \cdot \dvw{\rvec}{\rvec'}.
\label{eq:press2}
\end{equation}
It is important to observe that steady uniform flow makes no contribution whatsoever to the pressure in incompressible flow; the pressure in an unbounded flow is entirely due to vorticity\footnote{If the uniform flow is unsteady, then its rate of change contributes to pressure. Also, in a slightly compressible flow, the uniform flow would affect the pressure at large distance by modifying the travel time and directivity of acoustic waves between the vorticity and the observer \citep{powell:1}.}. The coupling between $\Vinf$ and $\vw$ that arises in the dynamic pressure is canceled by the modulation by $\Vinf$ inside the integral. Thus, only the modulation of vorticity by velocity induced by other vortex elements matters to pressure.

Equation~\eqref{eq:press2} shows that vorticity has a quadratic effect on pressure. To reveal this effect more clearly, we replace $\vw$ in \eqref{eq:press2} by the Biot-Savart integral \eqref{eq:vbiot}, obtaining
\begin{equation}
\press(\rvec) - \press_{\infty} = -\frac{1}{2} \rho \int_{\rvec'}\int_{\rvec''} \dvw{\rvec}{\rvec'}\cdot \dvw{\rvec}{\rvec''} + \rho \int_{\rvec'}\int_{\rvec''} \dvw{\rvec}{\rvec'}\cdot \dvw{\rvec'}{\rvec''}.
\end{equation}  
This form reveals an essentially triadic relationship between vorticity and pressure, illustrated in Figure~\ref{fig:triadic}: the pressure at $\rvec$ comprises a double sum of elementary interactions between vorticity at $\rvec'$ and $\rvec''$. Interestingly, a consequence of this relationship is that the pressure is invariant to a change of sign of the entire vorticity field. 

\subsection{Pressure from point vortices in the plane}

To illustrate this triadic interaction in a simple setting, let us consider a two-dimensional vorticity field consisting of two point vortices,
\begin{equation}
\w(\rvec) = \ez\left(\Gamma_{1} \delta(\rvec - \rvec_{1}) + \Gamma_{2} \delta(\rvec - \rvec_{2}) \right).
\end{equation}
The integrals can be evaluated exactly by virtue of the properties of the Dirac delta function, and in this two-dimensional setting, $\nabla \green(\rvec) = -\rvec/(2\pi |\rvec|^{2})$. The velocity induced by vortex $\vjdex$ is $\v_{\vjdex}(\rvec) = \nabla\green(\rvec-\rvec_{\vjdex}) \times \Gamma_{\vjdex} \ez$, and the resulting pressure field can be written as
\begin{equation}
\press(\rvec) - \press_{\infty} = -\frac{1}{2}\rho \Gamma^{2}_{1} \vorkern{\rvec}{\rvec_{1}}  - \frac{1}{2} \rho \Gamma^{2}_{2} \vorkern{\rvec}{\rvec_{2}} - \rho \Gamma_{1} \Gamma_{2} \vintkern{\rvec}{\rvec_{1}}{\rvec_{2}},
\label{eq:pvpress}
\end{equation}
where we have defined a {\em direct vortex kernel} for vortex $\vjdex$,
\begin{equation}
\vorkernbase(\rvec) \equiv \left| \nabla \green(\rvec) \right|^{2}  = \frac{1}{4\pi^{2}|\rvec|^{2}},
\label{eq:dirkern}
\end{equation}
and a {\em vortex interaction kernel},
\begin{align}
\vintkernbase(\rvec,\rvec') &= \vintkernbase^{(1)}(\rvec,\rvec') + \vintkernbase^{(2)}(\rvec,\rvec') \nonumber\\
&\equiv \nabla \green(\rvec) \cdot \nabla \green(\rvec') + \nabla\green(\rvec-\rvec')\cdot \left(\nabla\green(\rvec) - \nabla\green(\rvec')\right),
\label{eq:intkern}
\end{align}
which we have split into additive parts arising from the dynamic pressure term and the Lamb vector term, respectively. In the expression for $\vintkernbase^{(2)}$, we have used the fact that the Green's function's gradient is skew-symmetric: $\nabla\green(\rvec-\rvec') = -\nabla\green(\rvec'-\rvec)$.
 
\begin{figure}
\includegraphics{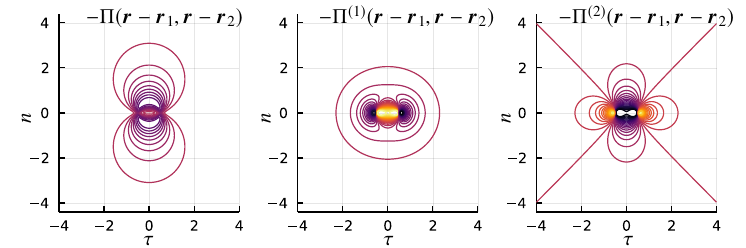}
\caption{Vortex interaction kernel for a pair of unit-strength point vortices at $\rvec_{1} = (-1/2,0)$ and $\rvec_{2} = (1/2,0)$. The two right panels depict the additive parts of the overall kernel. Each of the plots uses 41 contours between $-0.1$ and 0.1; blue is negative.} \label{fig:pvp1}
\end{figure}

There are a few notable features of expression \eqref{eq:pvpress}. First of all, each vortex makes an independent contribution to pressure via its direct vortex kernel. This direct contribution to the pressure field is always negative, regardless of the sign of the vortex, and is radially symmetric about the center of the vortex. These direct contributions are modified by the vortex interaction kernel, in a term that introduces the signs of the individual vortices into the pressure field. This kernel, $\vintkern{\rvec}{\rvec_{\vjdex}}{\rvec_{\vkdex}}$, is dependent only on the relative positions of the observation point $\rvec$ from each of the two vortex positions, $\rvec_{\vjdex}$ and $\rvec_{\vkdex}$. It is symmetric with respect to the members of the pair, $\vjdex$ and $\vkdex$, as is apparent from Figure~\ref{fig:pvp1}, which shows the kernel and its two additive parts.

The vortex interaction kernel is centered midway between the pair at $\rvecjk = (\rvec_{\vkdex}+\rvec_{\vjdex})/2$ and has directivity as indicated in the left panel of Figure~\ref{fig:pvp1}. It is apparent that the interaction kernel has much less influence along the pair's axis (the $\tau$ direction); its primary influence is perpendicular to this line, in the $n$ direction. We can write the kernel exactly as
\begin{equation}
\vintkern{\rvec}{\rvec_{\vjdex}}{\rvec_{\vkdex}} = \frac{1}{2\pi^{2}} \frac{|\axisjk\times(\rvec-\rvecjk))|^{2}}{|\rvec-\rvec_{\vjdex}|^{2}|\rvec-\rvec_{\vkdex}|^{2}},
\label{eq:vint0}
\end{equation}
where we denote the unit vector from vortex $\vjdex$ to vortex $\vkdex$ by $\axisjk = (\rvec_{\vkdex} - \rvec_{\vjdex})/\djk$, and $\djk = |\rvec_{\vkdex} - \rvec_{\vjdex}|$ the distance between the vortices. This form for the interaction kernel clearly shows that it, like the direct kernel, is purely positive. It is for this reason that we have explicitly pulled the negative sign out of equation \eqref{eq:pvpress}, to more clearly reveal the dependence of the sign of pressure on the vortex strengths.

Alternatively, we can write the kernel in a manner that emphasizes its scaling and directivity, as
\begin{equation}
\vintkern{\rvec}{\rvec_{\vjdex}}{\rvec_{\vkdex}} = \frac{1}{2\pi^{2} \djk^{2}} \frac{|\axisjk\times\rvecscale|^{2}}{|\rvecscale-\frac{1}{2}\axisjk|^{2}|\rvecscale+\frac{1}{2}\axisjk|^{2}},
\label{eq:vint}
\end{equation}
where by $\rvecscale = (\rvec - \rvecjk)/\djk$ we denote the vector from the vortex center to the observation point, re-scaled by $\djk$. This form distinguishes the scaling due to the pair's separation distance (the $1/\djk^{2}$ factor) from the scale-invariant directivity pattern. It also makes it easy to identify the far-field behavior, $|\rvecscale| \gg 1$, or $|\rvec - \rvecjk| \gg \djk$,
\begin{equation}
\vintkern{\rvec}{\rvec_{\vjdex}}{\rvec_{\vkdex}} \sim \frac{1}{2\pi^{2}|\rvec-\overline{\rvec}_{\vjdex\vkdex}|^{2}} \left| \axisjk\times\frac{\rvec-\overline{\rvec}_{\vjdex\vkdex}}{|\rvec-\overline{\rvec}_{\vjdex\vkdex}|}\right|^{2}.
\end{equation}  
Thus, the vortex interaction kernel has the same geometric decay, $1/r^{2}$, as that of the direct contributions and cannot generally be ignored. However, as \eqref{eq:vint} shows, the vortex interaction kernel's contribution becomes weaker in the near field, or equivalently, as the pair becomes further separated. 

\subsection{Systems of vortices}

It is a simple matter to extend the example of two point vortices to a larger set of $\nv$ point vortices. Using the same notation, the pressure field for this set is
\begin{equation}
\press(\rvec) - \press_{\infty} = -\frac{1}{2}\rho \sum_{\vjdex=1}^{\nv} \Gamma^{2}_{\vjdex} \vorkern{\rvec}{\rvec_{\vjdex}} - \rho \sum_{\vjdex=1}^{\nv} \sum_{\vkdex = 1}^{\vjdex-1} \Gamma_{\vjdex}\Gamma_{\vkdex} \vintkern{\rvec}{\rvec_{\vjdex}}{\rvec_{\vkdex}}.
\label{eq:pvpfull}
\end{equation}
The first term provides the direct contributions from each vortex; the second term comprises a sum over all of the $\nv(\nv+1)/2$ unique pairs in the set, using the vortex interaction kernel defined earlier. We can write this as a quadratic form in the vector of vortex strengths, $\Gamvec = [\Gamma_{1}\,\, \Gamma_{2}\cdots \Gamma_{\nv}]^{T}$:
\begin{equation}
\press(\rvec) - \press_{\infty} = -\frac{1}{2} \rho \Gamvec^{T} \matkern(\rvec)\Gamvec,
\end{equation}
where the elements of the symmetric positive semi-definite matrix $\matkern(\rvec)$ are
\begin{equation}
\matkernbase_{\vjdex\vkdex}(\rvec) = \left\{ \begin{array}{l} \vorkern{\rvec}{\rvec_{\vjdex}},\quad \vkdex = \vjdex, \\ \vintkern{\rvec}{\rvec_{\vjdex}}{\rvec_{\vkdex}}, \quad \vkdex \neq \vjdex.
\end{array}\right.
\end{equation}

\subsection{Regularized point vortices}
\label{sec:regularized}

In practice, we often rely on regularized point vortex elements rather than singular elements. There are two interpretations of the regularization process when viewed from the infinitesimal contribution \eqref{eq:dvw}. In one, we view the vorticity field as comprising a set of smooth blobs of vorticity in place of singular distribution. In the other, we still interpret the vorticity as a set of singular elements, but the Green's function and its gradient are replaced by versions that are convolved with a smooth regularization kernel. This convolution process is ultimately still present in the blob interpretation, after the infinitesimal contributions to velocity are integrated over space, so the interpretations both result in the same velocity field. 

A common choice of regularization kernel---and the one we use in this work---is the smooth algebraic form,
\begin{equation}
\tilde{\delta}_{\blobrad}(\rvec) = \frac{1}{\pi (|\rvec|^{2} + \blobrad^{2})^{2}},
\end{equation}
where $\blobrad$ is the blob radius. From this kernel, it can be shown that the regularized Green's function gradient is
\begin{equation}
\nabla \green_{\blobrad}(\rvec) = -\frac{\rvec}{2\pi \left(|\rvec|^{2} + \blobrad^{2}\right)}.
\label{eq:regdG}
\end{equation}
It is important to note that regularized vortices are approximate solutions to the incompressible Euler's equations \citep{hald:1j}, and as such, the pressure is determined in the same way as for singular point vortices. To determine this pressure, we utilize the second interpretation above, in which the vortices are still regarded as singular but the velocity they induce is regularized. As a result, we simply replace all instances of $\nabla\green$ in the pressure kernels \eqref{eq:dirkern} and \eqref{eq:intkern} with the regularized version \eqref{eq:regdG}, leading to regularized versions of these kernels that we denote by $\vorkernbase_{\blobrad}$ and $\vintkernbase_{\blobrad}$, respectively.


\subsection{Grid-based vorticity}
\label{sec:grid}

\begin{figure}
\begin{center}
\includegraphics[width=0.85\textwidth]{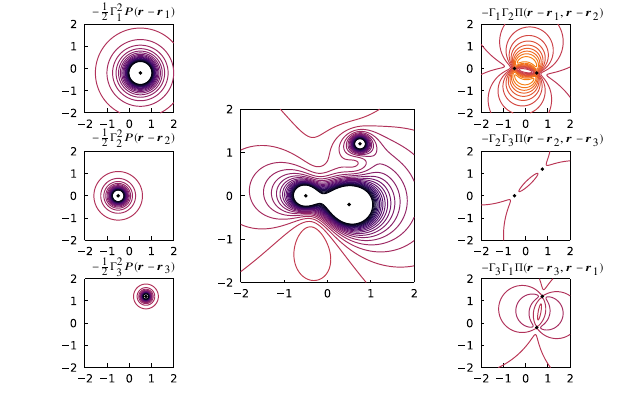}
\caption{Center: full pressure field from three Gaussian-distributed vortices (each with blob radius 0.1) at $\rvec_{1} = (0.5,-0.2)$ with strength $\Gamma_{1} = -1$; at $\rvec_{2} = (-0.5,0)$ and strength $\Gamma_{2} =0.5$; and at $\rvec_{3} = (0.75,1.2)$ with strength $\Gamma_{3} =-0.25$. Left panels depict, from top to bottom, the direct kernel for vortices 1, 2, and 3. Right panels show the interaction kernel for each pair of vortices. Vortex centers are depicted as black dots. Each of the plots uses 41 contours between $-0.05$ and 0.05; blue is negative.}\label{fig:pvp3}
\end{center}
\end{figure}

Finally, we should emphasize that, although we have derived the equations in this section for a finite system of point vortices, the forms of these relations also hold for a grid-based representation of a continuous (and unbounded) two-dimensional vorticity field, as in a computational fluid dynamics (CFD) simulation. In that context, the sums are carried out over all grid points with non-negligible vorticity: each vortex position represents the location of a grid point and its corresponding strength is replaced by the vorticity at that point, multiplied by the area of the surrounding grid cell. An example of a three-vortex system, whose pressure is computed on a Cartesian grid, is shown in Figure~\ref{fig:pvp3}. To emphasize that the form extends to distributions of vorticity and not just isolated point vortices, this example depicts Gaussian-distributed vortices. The overall pressure field in the center panel is the sum of the fields in all of the satellite panels. The satellite panels on the right exhibit the same interaction kernel---shifted, rotated, and re-scaled for each pair.

\section{The expectation of vorticity under a Gaussian mixture model for vortex states}
\label{sec:expect-vort}

In this section, we derive the expectation of the vorticity field (\ref{eq:expected-vort}) when the vortex states are described by a Gaussian mixture model $\approxprob(\state)$ given by (\ref{eq:gmm}). (We omit the subscript $\truemeas$ from this probability here for brevity.) We start with the singular vorticity representation (\ref{eq:sing-vort}), and seek to evaluate the integral
\begin{equation}
\expect_{\approxprob}[\w](\rvec) = \int \w(\rvec, \state) \approxprob(\state)\,\mathrm{d}\state,
\label{eq:expect-vort-def}
\end{equation}
where the notation $\w(\rvec, \state)$ explicitly represents the dependence of the singular vortex field on the state vector $\state$. It is sufficient to consider the integration of just a single Gaussian component, since the overall expectation will simply be a linear combination of the $K$ components. Thus, we seek the integral
\begin{equation}
\expect_{\normaldist}[\w](\rvec) = \int \w(\rvec; \state) \normaldist(\state|\gmmmeanbase,\gmmcovarbase) \,\mathrm{d}\state = \sum_{\vjdex=1}^{\nv} \int \Gamma_{\vjdex} \delta(\rvec-\rvec_{\vjdex}) \normaldist(\state|\gmmmeanbase,\gmmcovarbase)\,\mathrm{d}\state.
\label{eq:expect-gauss-vort}
\end{equation}
Let us recall that the state and covariance are organized as shown in (\ref{eq:stateblock}) and (\ref{eq:covarblock}), respectively. The integration over $\state$ is multiplicatively decomposable into integrals over the states of the individual vortex elements, $\mathrm{d}\state = \mathrm{d}\state_{1}\mathrm{d}\state_{2}\cdots\mathrm{d}\state_{\nv}$. When the integral in (\ref{eq:expect-gauss-vort}) for vortex $\vjdex$ in the sum is carried out, the integrals over all states of the other other vortices $\videx \neq \vjdex$ represent marginalizations of the probability distribution over these vortices. Using properties of Gaussians \citep{bishopbook}, it is easy to show that this marginalized distribution is simply a Gaussian distribution over the states of vortex $\vjdex$,
 \begin{equation}
\expect_{\normaldist}[\w](\rvec) = \sum_{\vjdex=1}^{\nv} \int \Gamma_{\vjdex} \delta(\rvec-\rvec_{\vjdex}) \normaldist(\state_{\vjdex}|\gmmmeanbase_{\vjdex},\gmmcovarbase_{\vjdex\vjdex})\,\mathrm{d}\state_{\vjdex}.
\label{eq:expect-gauss-vort-j}
\end{equation}
Now, we can decompose the integral into the individual states of vortex $\vjdex$, $\mathrm{d}\state_{\vjdex} = \mathrm{d}\rvec_{\vjdex}\mathrm{d}\Gamma_{\vjdex}$, and the state $\state_{\vjdex}$ and covariance $\gmmcovarbase_{\vjdex\vjdex}$ partitioned accordingly, as in their definitions (\ref{eq:stateblockj}) and (\ref{eq:covarblockjk}). To assist the calculations that follow, it is useful to write the joint probability distribution for the strength and position $\approxprob(\Gamma_{\vjdex},\rvec_{\vjdex}) = \normaldist(\state_{\vjdex}|\gmmmeanbase_{\vjdex},\gmmcovarbase_{\vjdex\vjdex})$ in the conditional form $\probdist(\Gamma_{\vjdex},\rvec_{\vjdex}) = \probdist(\Gamma_{\vjdex}|\rvec_{\vjdex}) \probdist(\rvec_{\vjdex})$, where
\begin{equation}
\probdist(\rvec_{\vjdex}) = \normaldist(\rvec_{\vjdex}|\overline{\rvec}_{\vjdex},\covar_{\rvec_{\vjdex} \rvec_{\vjdex}}).
\end{equation}
Again, using properties of Gaussians, the conditional probability $\probdist(\Gamma_{\vjdex}|\rvec_{\vjdex})$ can be shown (by completing the square) to be
\begin{equation}
\probdist(\Gamma_{\vjdex}|\rvec_{\vjdex}) = \normaldist(\Gamma_{\vjdex}|\mu_{\Gamma_{\vjdex}|\rvec_{\vjdex}},\covar_{\Gamma_{\vjdex}|\rvec_{\vjdex}}),
\end{equation}
where the mean and covariance are, respectively,
\begin{equation}
\mu_{\Gamma_{\vjdex}|\rvec_{\vjdex}} = \overline{\Gamma}_{\vjdex} + \covar_{\Gamma_{\vjdex} \rvec_{\vjdex}} \covar_{\rvec_{\vjdex}\rvec_{\vjdex}}^{-1} (\rvec_{\vjdex} - \overline{\rvec}_{\vjdex}), \qquad \covar_{\Gamma_{\vjdex}|\rvec_{\vjdex}} = \covar_{\Gamma_{\vjdex}\Gamma_{\vjdex}} - \covar_{\Gamma_{\vjdex} \rvec_{\vjdex}} \covar_{\rvec_{\vjdex}\rvec_{\vjdex}}^{-1} \covar_{\rvec_{\vjdex}\Gamma_{\vjdex}}.
\end{equation}

In this partitioned form, we can evaluate the integrals over $\rvec_{\vjdex}$ and $\Gamma_{\vjdex}$ in (\ref{eq:expect-gauss-vort-j}). The integral over $\rvec_{\vjdex}$ is particularly easy to evaluate because of the properties of the Dirac delta function. As a result, $\rvec_{\vjdex}$ is replaced everywhere by the observation point $\rvec$. We are thus left with the integral
\begin{equation}
\expect_{\normaldist}[\w](\rvec) = \sum_{\vjdex=1}^{\nv} \normaldist(\rvec|\overline{\rvec}_{\vjdex},\covar_{\rvec_{\vjdex} \rvec_{\vjdex}}) \int \Gamma_{\vjdex}  \normaldist(\Gamma_{\vjdex}|\mu_{\Gamma_{\vjdex}|\rvec_{\vjdex}},\covar_{\Gamma_{\vjdex}|\rvec_{\vjdex}})\,\mathrm{d}\Gamma_{\vjdex},
\label{eq:expect-gauss-vort-j2}
\end{equation}
where $\rvec_{\vjdex}$ is replaced by $\rvec$ in the mean, $\mu_{\Gamma_{\vjdex}|\rvec_{\vjdex}}$. This final integral is simply the expectation of $\Gamma_{\vjdex}$ over the conditional distribution, and its value for a Gaussian is the mean, $\mu_{\Gamma_{\vjdex}|\rvec_{\vjdex}}$. Thus, we arrive at
\begin{equation}
\expect_{\normaldist}[\w](\rvec) = \sum_{\vjdex=1}^{\nv} \left[ \overline{\Gamma}_{\vjdex} + \covar_{\Gamma_{\vjdex} \rvec_{\vjdex}} \covar_{\rvec_{\vjdex}\rvec_{\vjdex}}^{-1} (\rvec - \overline{\rvec}_{\vjdex})\right] \normaldist(\rvec|\overline{\rvec}_{\vjdex},\covar_{\rvec_{\vjdex} \rvec_{\vjdex}}) 
\label{eq:expect-gauss-vort-jfinal}
\end{equation} 
The final result (\ref{eq:expected-vort}) follows easily by introducing (\ref{eq:expect-gauss-vort-jfinal}) into the mixture model.

\section{Linearization of the observation operator in Bayes theorem}
\label{sec:bayes-gauss}

Let us suppose that the prior is Gaussian rather than uniform, with mean $\state_{0}$ and covariance $\covar_{0}$. (Ultimately, we will allow this covariance to become infinitely large.) Thus,
\begin{equation}
\probdist_{0}(\state) = \normaldist(\state|\state_{0},\covar_{0}). 
\end{equation}
The likelihood is also assumed Gaussian about the observation prediction $\observe(\state)$ with covariance $\noisecovar$, as in \eqref{eq:likenormal}, but now we will linearize the observation operator about the true state $\truestate$, as in \eqref{eq:taylor}. This can be written as
\begin{equation}
    \observe(\state) \approx \observemat \state + b,
\end{equation}
where $\observemat \equiv \nabla\observe(\truestate)$ is the Jacobian of the observation at the true state, and $b = \observe(\truestate) - \observemat\truestate$.

With Gaussian prior and likelihood and a linear relationship between $\meas$ and $\state$, the joint distribution over these variables is also Gaussian. We can make use of the properties of multivariate Gaussians to obtain all of well-known results that follow; the reader is referred to \cite{bishopbook} for more details. The mean and covariance of the joint variable $\joint = (\state,\meas)$ are, respectively,
\begin{equation}
\mu_{\State,\Meas} = \begin{pmatrix}
    \state_{0} \\ \observemat\state_{0} + b
\end{pmatrix}
\end{equation}
and
\begin{equation}
    \covar_{\State,\Meas} = \begin{pmatrix}
        \covar_{0} & \covar_{0} \observemat^T \\ \observemat \covar_{0} & \noisecovar + \observemat \covar_{0} \observemat^T
    \end{pmatrix}.
\end{equation}

To obtain the Gaussian form of the posterior distribution, we seek the mean and covariance of the conditional $\probdist(\state|\meas)$, which is obtained by starting from the log of the joint distribution
\begin{equation}
-\frac{1}{2} (\joint - \mu_{\State,\Meas})^T\covar^{-1}_{\State,\Meas} (\joint - \mu_{\State,\Meas})
\end{equation}
and rewriting it as a quadratic form in $\state$ only, with $\meas$ set equal to the true observation, $\truemeas$, and assumed known. Again, using well-known identities involving the inverse of a partitioned matrix, we arrive at the conditional mean and covariance,
\begin{equation}
    \mu_{\State|\truemeas} = (\covar_{0}^{-1} + \observemat^T \noisecovar^{-1} \observemat)^{-1} \left( \observemat^T \noisecovar^{-1} (\truemeas - b) + \covar_{0}^{-1} \state_{0}\right)
\end{equation}
and
\begin{equation}
    \covar_{\State|\truemeas} = (\covar_{0}^{-1} + \observemat^T \noisecovar^{-1} \observemat)^{-1}. 
\end{equation}
These results balance the prior mean and covariance with the information gained from the observation, $\truemeas$. However, if the prior covariance grows to infinity, $\covar_{0}^{-1} \rightarrow 0$, reflecting our lack of prior knowledge, then all dependence on the prior vanishes, and we end up with
\begin{equation}
    \mu_{\State|\truemeas} = \truestate + (\observemat^T \noisecovar^{-1} \observemat)^{-1} \observemat^T \noisecovar^{-1} \left(\truemeas - \observe(\truestate) \right).
    \label{eq:cond-mean-zeroprior}
\end{equation}
and
\begin{equation}
    \covar_{\State|\truemeas} = (\observemat^T \noisecovar^{-1} \observemat)^{-1},
    \label{eq:cond-covar-zeroprior}
\end{equation}
in which we have also substituted the specific form of $b$ in our linearized model and simplified. The second term in the mean represents a bias error that arises when the true observation differs from the model evaluated at the true state, as from the error $\noise^*$ in a single realization of the measurements.

\bibliographystyle{jfm}

\end{document}